\pdfoutput=1
\documentclass[aps,twocolumn,floatfix,prb]{revtex4}
\usepackage{graphicx,amsmath,bbm}

\newcommand{\vk}{\vec k}
\newcommand{\ve}{\vec e}

\newcommand{\vq}{\vec q}

\renewcommand{\vr}{\vec r}

\newcommand{\vs}{\vec s}

\newcommand{\vsigma}{\mbox{\boldmath $\sigma$}}
\newcommand{\vrho}{\mbox{\boldmath $\rho$}}
\newcommand{\vrhosuper}{\mbox{\scriptsize \boldmath $\rho$}}

\renewcommand{\Im}{\mbox{Im}\,}
\renewcommand{\Re}{\mbox{Re}\,}
\renewcommand{\vec}[1]{\mathbf{#1}}

\newcommand{\ua}{\uparrow}
\newcommand{\da}{\downarrow}

\begin{document}

\title{Dephasing in ferromagnetic nanowires: The role of spin waves}
\date{\today}

\author{J.\ Danon}
\author{P.~W.\ Brouwer}
\affiliation{Dahlem Center for Complex Quantum Systems and Fachbereich Physik, Freie Universit\"{a}t Berlin, Arnimallee 14, 14195 Berlin, Germany}

\begin{abstract}
We present a calculation of the dephasing time of electrons in a ferromagnet relevant for the conductance fluctuations. We focus on the contribution from the interaction with spin waves. Explicit results are presented for a quasi-one-dimensional systems. Going beyond previous calculations, we do not restrict ourselves to the limit of a small exchange splitting compared to the electronic elastic scattering time, nor does our calculation rely on the diffusion approximation to describe electronic transport.
\end{abstract}

\maketitle

\section{Introduction}
Effects of quantum coherence on transport properties of disordered metals have been intensively investigated in the past few decades, and are by now well understood.\cite{imry:book,akkermans} Examples of interference phenomena are weak localization\cite{bergmann19841,aa:book,RevModPhys.57.287} and Aharonov-Bohm oscillations.\cite{buttiker:physlett,intel} Another signature of quantum coherent transport is the phenomenon of `universal conductance fluctuations':\cite{PhysRevLett.55.1622,altshuler1985,PhysRevB.35.1039} The conductance of a disordered metal exhibits random but reproducible fluctuations as a function of an external control parameter, such as the magnetic field or a gate voltage. The typical amplitude of these fluctuations is of order $e^2/h$ at zero temperature, independent of the microscopic details of the disorder. At finite temperature, the conductance fluctuations are suppressed because of thermal averaging and dephasing.

Quantum coherent transport in {\it ferromagnetic} disordered metals has attracted attention only at a later stage, stimulated by the discovery of the giant magnetoresistance\cite{gmr1,gmr2} and by the emergence of the field of spintronics.\cite{spinrev} In spite of the presence of an internal magnetic field, weak localization is observed in ferromagnetic conductors if the sample size or phase-breaking length is small enough that the effect of the orbital magnetic field or spin-orbit scattering can be neglected on that length scale.\cite{kn:aprili1997,kn:wei2006} Several groups have measured conductance fluctuations in ferromagnets\cite{kn:kasai2003,kn:lee2007,kn:wei2006,kn:vila2007,kn:lee2004,kn:bolotin2006,kn:neumaier2008} and obtained estimates for the dephasing rates and their temperature dependence from these.\cite{kn:kasai2003,kn:lee2007,kn:wei2006,kn:vila2007,kn:neumaier2008} In some of these experiments, the measured dephasing rates were significantly larger than in otherwise comparable normal metals,\cite{kn:wei2006,kn:kasai2003,kn:lee2007} but the microscopic mechanism responsible for this enhanced dephasing could not be identified.


With a static and spatially uniform magnetization, the propagation of majority electrons and minority electrons in the ferromagnetic metal is decoupled, and the quantum corrections to transport are essentially equal to those in a normal metal. Differences only appear in the presence of spin-orbit coupling, or when the magnetization is not constant as a function of time or position, as is the case with domain walls or in the presence of spin waves (magnons). The weak localization correction to the conductivity of ferromagnets has been theoretically investigated in the presence of spin-orbit interaction,\cite{PhysRevB.64.144423} as well as in the presence of domain walls.\cite{takane2} The effect of domain walls or spin-orbit coupling on the conductance fluctuations has also been explored.~\cite{PhysRevLett.78.3773,PhysRevLett.81.3215,kn:adam2006b} 

Spin waves, fluctuations of the magnetization direction that vary in time and space, are expected to contribute to the dephasing of electrons in a ferromagnet. Takane has evaluated the dephasing rate relevant for the conductance fluctuations in quasi-one-dimensional wires.\cite{JPSJ.72.1155} Muttalib and W\"{o}lfle performed a similar calculation for the phase relaxation rate in thin ferromagnetic films.\cite{muttalibwoelfle} The calculation of Ref.\ \onlinecite{JPSJ.72.1155}
 is restricted to the limit where the exchange splitting $\Delta$ is much smaller than the electronic elastic scattering rate $\hbar/\tau_{\rm el}$, which is typically not the case in (elemental) ferromagnets. Reference \onlinecite{muttalibwoelfle} addresses both the clean and diffusive limits, but still relies on the inequality $q_T l_{\rm el} \ll 1$, where $q_T$ is the wave number of a thermal spin wave, and $l_{\rm el} = v_{\rm F} \tau_{\rm el}$ the elastic mean free path. This limits the applicability of the theory to rather low temperatures, where dephasing from electron-electron interactions is likely to dominate over spin-wave-induced dephasing. 

In this work we present a calculation of the spin-wave-induced dephasing rate $1/\tau_{\phi}$ of electrons in a ferromagnet. We derive a general expression for $1/\tau_{\phi}$ in terms of the spin wave dispersion relation for a model of Gaussian-white-noise disorder, and calculate the explicit temperature dependence for the special case of a wire geometry. In contrast to previous calculations, we do not restrict ourselves to the limit of small $\Delta\tau_{\rm el}/\hbar$ or employ the diffusion approximation. We not only find qualitatively different results for the clean limit $\Delta\tau_{\rm el}/\hbar \gg 1$ and dirty limit $\Delta\tau_{\rm el}/\hbar \ll 1$, but inside the clean limit we also find qualitatively different dephasing rates for the cases when majority and minority electrons have equal or different elastic scattering rates. In the most realistic regime $\Delta\tau_{\rm el}/\hbar \gg 1$, we find that the temperature-dependent dephasing rate obeys a power law: at low temperatures $1/\tau_{\phi} \propto T^{5/2}$, whereas at higher temperatures $1/\tau_{\phi} \propto T^{3/2}$. These temperature dependences are such that dephasing from electron-electron interactions, which scales $\propto T^{2/3}$ in a wire geometry,\cite{aa:book} is stronger than dephasing from spin waves at the lowest temperatures, but they leave open the possibility that spin waves are the dominant source of dephasing at higher temperatures.

The structure of the paper is as follows. In Sec.\ \ref{sec:2} we first present all ingredients of the model we use to describe the propagation of $s$-band electrons in the disordered ferromagnet and their interaction with the localized spins of $d$-band electrons. In Sec.\ \ref{sec:3} we describe a calculation of the dephasing rate in a semi-classical picture, considering the coherent propagation of an electron in a fluctuating classical exchange field. In Sec.\ \ref{sec:4} we then outline our full diagrammatic calculation. Details are kept for the Appendix. In Sec.\ \ref{sec:5} we finally calculate the temperature dependence of the dephasing rate, focusing for simplicity on a quasi-one-dimensional wire geometry. We conclude with a discussion of our results in Sec.\ \ref{sec:6}.

\section{Model}
\label{sec:2}

For the conduction electrons in the disordered ferromagnet, we consider the effective single-particle Hamiltonian
\begin{equation}
  H = H_{\rm kin} + V + H_{sd},
\label{eq:ham1}
\end{equation}
where $H_{\rm kin} = p^2/2m$ is the kinetic energy, $V$ the impurity potential, and $H_{sd}$ describes the exchange between the conduction electrons and the $d$ electron spins. For $H_{sd}$ we take the simple model
\begin{equation}
  H_{sd} = -J \vs(\vr) \cdot \vsigma,
  \label{eq:hsd}
\end{equation}
where $J$ is the exchange constant and $\vs(\vr)$ is the spin density of the $d$-electrons expressed in units of $\hbar$. The exchange constant therefore has dimensions energy times volume. In a ferromagnet, the magnetization $\vs(\vr)$ fluctuates around a nonzero mean value $\overline{\vs} = \overline{s} \ve_s$, which we take along the $z$ axis. To linear order in the fluctuations, only transverse fluctuations of $\vs$ need to be considered, so that we can write $H_{sd}$ as
\begin{equation}
  H_{sd} = - \frac{1}{2} \Delta \sigma_z +
  H_{sd,\perp}, \label{eq:hamsd}
\end{equation}
where $\Delta = 2 J \overline{s}$ is the exchange splitting between majority and minority electrons and
\begin{equation}
  H_{sd,\perp}({\bf r},t) = 
  J \left( \begin{array}{cc} 0 & s_-({\bf r},t) \\ s_+({\bf r},t) & 0 \end{array} \right),
\end{equation}
where $s_{\pm} = s_x \pm i s_y$. (The exchange splitting $\Delta$ may be different from $J \overline{s}$ if exchange interactions between conduction electrons are taken into account. Here and below, we will not make use of the relation $\Delta = 2J \overline{s}$ implied by our model, but consider $\Delta$ an independent parameter instead.)

In our calculations, we will take $\Delta$ small in comparison to the Fermi energy, so that majority and minority electrons have the same density of states $\nu$ at the Fermi level and the same Fermi velocity $v_{\rm F}$. In order to model the effect that majority and minority electrons have different scattering rates in realistic ferromagnets, we take different disorder potentials for majority and minority electrons,
\begin{equation}
  V = \left( \begin{array}{cc} V_{\ua} & 0 \\ 0 & V_{\da} \end{array} \right).
\end{equation}
Here $V_{\ua}$ and $V_{\da}$ are Gaussian white noise potentials with correlation functions
\begin{eqnarray}
  \langle V_{\ua}(\vr) V_{\ua}(\vr') \rangle &=&
  \frac{1}{2 \pi \nu \tau_{\ua}} \delta(\vr-\vr'), \nonumber \\
  \langle V_{\da}(\vr) V_{\da}(\vr') \rangle &=&
  \frac{1}{2 \pi \nu \tau_{\da}} \delta(\vr-\vr'), \\
  \langle V_{\ua}(\vr) V_{\da}(\vr') \rangle &=&
  \frac{1}{2 \pi \nu}
  \left( \frac{1}{2\tau_{\ua}} + \frac{1}{2\tau_{\da}} 
  - \frac{1}{\tau_{\rm d}} \right) \delta(\vr-\vr'), \nonumber
\end{eqnarray}
where $\tau_{\ua}$ and $\tau_{\da}$ are the elastic mean free times for majority and minority electrons, respectively, and $\tau_{\rm d}$ is a time that describes the degree of correlation between the effective impurity potentials for majority and minority electrons. The case of equal impurity potentials for majority and minority electrons corresponds to the case $\tau_{\ua} = \tau_{\da}$ and $\tau_{\rm d} \to \infty$.

The alternative to our implementation of different scattering rates for majority and minority electrons through a spin-dependent impurity potential is to take the Fermi energy difference between majority electrons and minority electrons seriously. However, taking $\Delta$ to be comparable to the Fermi energy $\varepsilon_{\rm F}$ is incompatible with the semiclassical limiting procedure at the foundation of the diagrammatic calculation of the dephasing time,\cite{altshuler1985} because it no longer allows one to separate diagrams that are small by a factor $\hbar/\varepsilon_{\rm F} \tau$ (which are usually neglected) from those that are small by a factor $\hbar/\Delta \tau$ (which need to be kept for a diagrammatic calculation of the dephasing rate). The choice of a spin-dependent impurity potential circumvents these problems.\cite{JPSJ.72.1155}

In second-quantized language, the transverse $sd$ exchange Hamiltonian $H_{sd,\perp}$ takes the form
\begin{equation}
  H_{sd,\perp} =  -\frac{J}{V} \sum_{\vk,\vq} \left[c^{\dagger}_{\vk+\vq,\ua} c_{\vk,\da}
  s_{\vq,-} + c^{\dagger}_{\vk+\vq,\da} c_{\vk,\ua} s_{\vq,+} \right],
\end{equation}
with $s_{\vq,\pm} = \int d\vr s_{\pm}(\vr , t) e^{-i \vq \cdot \vr}$ the Fourier transform of the spin density and $V$ the volume of the ferromagnet. Dynamical processes involving the excitation and absorption of a $d$-band spin wave are characterized by the susceptibility
\begin{equation}
 \chi^{\rm R}_{-+}(\vq,\tau) = -\frac{1}{V}i\Theta(\tau) \langle [s_{\vq,-}(\tau),s_{\vq,+}(0)] \rangle,
\label{eq:spincorr}
\end{equation}
where $\Theta(\tau) = 1$ for $\tau > 0$ and $\Theta(\tau) = 0$ otherwise is the Heaviside step function. The susceptibility $\chi^{\rm R}_{-+}(\vq,\tau)$ describes the response of the $d$-electron spin density to an applied magnetic field. The Fourier transform $\chi^{\rm R}_{-+}(\vq,\omega)$ is conveniently expressed in terms of the spin wave frequencies $\omega_{\vq}^{\rm sw}$,\cite{kubospinsus,JPSJ.72.1155}
\begin{eqnarray}
  \chi^{\rm R}_{-+}(\vq,\omega) &=&
  \int d\tau \chi^{\rm R}_{-+}(\vq,\tau) e^{i \omega \tau} \\ &=&
  - \frac{2 \overline{s}}{\omega - \omega_{\vq}^{\rm sw} + i\eta},
  \label{eq:chisw}
\end{eqnarray}
where $\eta$ is a positive infinitesimal.

\section{Semiclassical picture}
\label{sec:3}

The spin wave contribution to the dephasing rate in ferromagnetic metals can be understood from a semiclassical argument. Hereto we consider the $d$-electron spin density $\vs(\vr)$ as a classical variable, and look at the effect of fluctuations of $\vs(\vr)$ on the coherent propagation of conduction electrons. We treat the orbital degrees of freedom classically, with conduction electrons moving along classical trajectories $\vr(t)$. Their time evolution is governed by the $sd$-exchange Hamiltonian (\ref{eq:hsd}), which is taken to act on the spinor degrees of freedom of the conduction electrons only. In this approach, the Hamiltonian $H_{sd}$ depends on time $t$ explicitly through the time dependence of the $d$-electron spin density $\vs$ and implicitly through the time dependence of the conduction electron's position $\vr(t)$, combined with the $\vr$ dependence of $\vs$.

As in the previous section, we separate $H_{sd}$ into a (large) time-independent part $\overline{H}_{sd} = - (\Delta/2) \sigma_z$ and a (small) time-dependent part $H_{sd,\perp}$ describing the transverse fluctuations of the $d$-electron spin density.
We describe the time-evolution of a spin described by the Hamiltonian $H_{sd}$ in the interaction picture, in which $H_{sd,\perp}$ is treated as a perturbation. The time-evolution operator reads
\begin{equation}
  U^{(I)}(\tau;t) = \mathcal{T} e^{-\frac{i}{\hbar} \int_{t}^{t+\tau} dt' H_{sd,\perp}^{(I)}(t')},
\label{eq:timeevolpert}
\end{equation}
where the perturbation is written in the interaction picture, 
\begin{equation}
  H_{sd,\perp}^{(I)}(t) = e^{i t \Delta \sigma_z/2\hbar} H_{sd,\perp}(t) e^{-i t \Delta \sigma_z/2 \hbar},
\end{equation}
and $\mathcal{T}$ denotes the time-ordering operator. The relation between $U^{(I)}$ and the time-evolution operator in the Schr\"{o}dinger picture $U$ is 
\begin{equation}
  U(\tau;t) = e^{i (t+\tau) \Delta \sigma_z/2 \hbar} U^{(I)}(\tau;t) e^{-i t \Delta \sigma_z/2 \hbar}.
\end{equation}

We now calculate the average $\overline{U(\tau)}$ with respect to the time-dependent fluctuations of the transverse spin density $s_{\pm}$. In order to separate the explicit time dependence of $s_{\pm}(\vr[t],t)$ through the time argument $t$ and the implicit time dependence through the $t$ dependence of the position $\vr$, we consider a fixed trajectory, but at different starting times $t$. We note that the off-diagonal elements of the evolution operator vanish upon taking the time-average, because they contain an odd power of the transverse $d$-electron spin densities $s_{\pm}$, so that it is sufficient to consider the diagonal elements $\overline{U_{\ua\ua}(\tau)}$ and $\overline{U_{\da\da}(\tau)}$. We also note, that in the expansion of the the diagonal element $U_{\uparrow\uparrow}^{(I)}(\tau;t)$ the transverse spin densities always occur pairwise in the combination $s_-(\vr[t_2],t_2) s_+(\vr[t_1],t_1)$, with $t_2 > t_1$. For $\tau$ much longer than the time scale for fluctuations of the magnetization, these events appear well separated in time, and one finds that the leading contribution to $\overline{U_{I,\uparrow\uparrow}(\tau)}$ is given by
\begin{equation}
  \overline{U_{I,\uparrow\uparrow}(\tau)} =
  e^{ -(J/\hbar)^2 \int_{0}^{\tau} 
  d\tau_1 \int_{0}^{\tau} d\tau_2 F_{-+}(\tau_2,\tau_1)},
  \label{eq:UIup}
\end{equation}
where
\begin{eqnarray}
  F_{-+}(\tau_2,\tau_1) &=&   
  \overline{s_-(\vr[t+\tau_2],t+\tau_2) s_+(\vr[t+\tau_1],t+\tau_1)}
  \nonumber \\ && \mbox{} \times
  \Theta(\tau_2-\tau_1) e^{-i \Delta(\tau_2-\tau_1)/\hbar}.
  \label{eq:Fdefup}
\end{eqnarray}
The averaging bar $\overline{\cdots}$ denotes an average with respect to $t$. For long times, the magnitude of $\overline{U_{I,\uparrow\uparrow}(\tau)}$ decays $\propto \exp(-\tau/2\tau_{\phi,\uparrow})$, where $\tau_{\phi,\uparrow}$ is the dephasing time. From Eq.\ (\ref{eq:UIup}) we then conclude that
the dephasing time $\tau_{\phi,\uparrow}$ is given by
\begin{equation}
  \frac{1}{\tau_{\phi,\uparrow}} =
  2 \frac{J^2 }{\hbar^2}
  \Re 
  \int_0^{\infty} d\tau' \langle F_{-+}(\tau_1+\tau',\tau_1) \rangle_{\tau_1},
  \label{eq:tauphiup}
\end{equation}
where the brackets $\langle \cdots \rangle$ indicate an average along the trajectory $\vr(t+\tau_1)$. Repeating the same analysis for minority electrons gives
\begin{equation}
  \frac{1}{\tau_{\phi,\downarrow}} =
  2 \frac{J^2}{\hbar^2} \Re \int_0^{\infty} d\tau' \langle F_{+-}(\tau_1+\tau',\tau_1) \rangle_{\tau_1}, \label{eq:tauphidown}
\end{equation}
with
\begin{eqnarray}
  F_{+-}(\tau_2,\tau_1) &=&   
  \overline{s_+(\vr[t+\tau_2],t+\tau_2) s_-(\vr[t+\tau_1],t+\tau_1)}
  \nonumber \\ && \mbox{} \times
  \Theta(\tau_2-\tau_1) e^{i \Delta(\tau_2-\tau_1)/\hbar}.
  \label{eq:Fdefdown}
\end{eqnarray}

It remains to evaluate the time averages in Eqs.\ (\ref{eq:Fdefup}) and (\ref{eq:Fdefdown}) and to perform the average along the trajectory in Eqs.\ (\ref{eq:tauphiup}) and (\ref{eq:tauphidown}). The spin correlation function required for the calculation of $F_{-+}$ follows directly from the fluctuation-dissipation theorem,\cite{ll5}
\begin{eqnarray}
  \overline{s_-(\vr+\vrho,t+\tau) s_+(\vr,t)} &=&
  \int \frac{d\vq d\omega}{(2 \pi)^4} \frac{2T}{\omega} e^{i \vq \cdot \vrhosuper -i \omega \tau}  \nonumber \\ && \mbox{} \times
  \text{Im}\{\chi_{-+}(\omega,{\bf q})\},
\label{eq:corr}
\end{eqnarray}
where $\chi_{-+}$ is the magnetic susceptibility, which describes the magnetization response to an applied magnetic field (note that we have set for convenience $k_B = 1$). The imaginary part of the response function $\chi_{-+}(\omega,{\bf q})$ is determined by the spin wave spectrum, see Eq.\ (\ref{eq:chisw}),
\begin{equation}
  \Im \{\chi_{-+}(\omega,{\bf q})\} =
  2\pi \overline{s}
  \delta(\omega - \omega_{\vq}^{\rm sw}).
\end{equation}

\begin{figure}[t]
\begin{center}
\includegraphics[width=82mm]{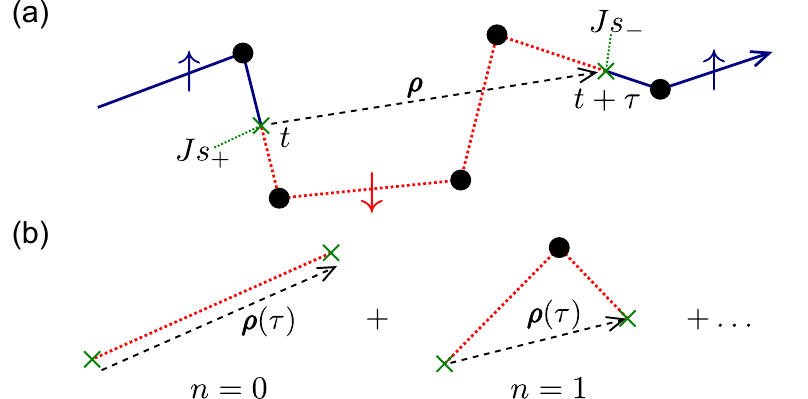}
\caption{(Color online) Illustration of our semiclassical calculation. (a) A spin-up electron propagates ballistically through the ferromagnet, its direction of propagation regularly being randomized by scattering off impurities (black circles). At time $t$, it interacts with the fluctuating field of the $d$-band electrons and its spin is flipped down. After a time $\tau$, the electron has traveled over a distance $\boldsymbol \rho$ and it interacts again with the $d$-band electrons, its spin being flipped up again. Since the interactions occur randomly in time and space, the phase of the electron becomes randomized while propagating. (b) The part of the trajectory important for dephasing is when the electron carries spin down. When averaging over the displacement $\boldsymbol\rho(\tau)$ during this part, we sum over trajectories with no scattering events ($n=0$), with one scattering event ($n=1$), etc.}
\label{fig:semclas}
\end{center}
\end{figure}
For the average along the trajectory $\vr(t)$ we consider $s$-wave impurity scattering with elastic mean free time $\tau_{\rm el} = \tau_{\ua} = \tau_{\da}$ and ballistic propagation between scattering events. In Fig.\ \ref{fig:semclas}a we illustrated a part of a trajectory for a spin-up electron: The electron scatters off impurities (black circles), and at times $t$ and $t+\tau$ it interacts with the fluctuating field of the $d$-band spins causing a spin flip. The electron thus carries spin down for a time $\tau$ (the red dotted part of the trajectory), and during this time it travels over a distance $\boldsymbol\rho$.

The average over trajectories in Eqs. (\ref{eq:tauphiup}) and (\ref{eq:tauphidown}) amounts to averaging $\langle e^{-i \vq\cdot\boldsymbol\rho (\tau) } \rangle_{\boldsymbol\rho}$ over all possible displacements $\boldsymbol\rho$ in time $\tau$. Summing over all possible number $n$ of scattering events in the time interval $\tau$ (see Fig.\ \ref{fig:semclas}b), we obtain
\begin{eqnarray}
  \frac{1}{\tau_{\phi}} &=&
  \int \frac{d\vq}{(2 \pi)^3}
  \frac{4 J^2 T \overline{s}}{\hbar^3\omega_{\vq}^{\rm sw}} \Re
  \sum_{n=0}^{\infty} \frac{1}{\tau_{\rm el}^{n}}
  \left( \frac{i K}{2 q v_{\rm F}} \right)^{n+1} \nonumber \\ &=&
  \int \frac{d\vq}{(2 \pi)^3}
  \frac{4 J^2 T \overline{s}}{\hbar^3\omega_{\vq}^{\rm sw}}
  \Re
  \frac{\tau_{\rm el} i K}{2 q v_{\rm F} \tau_{\rm el} - i K},
\end{eqnarray}
where we have dropped the spin index for $\tau_{\phi}$, because minority electrons and majority electrons have equal dephasing times in the semiclassical model, and 
\begin{eqnarray}
  K &=& -2 i q v_{\rm F}
  \int_0^{\infty} d\tau 
  e^{-i (\Delta/\hbar+\omega_{\vq}^{\rm sw}) \tau - \tau/\tau_{\rm el}} 
  \nonumber \\ && \mbox{} \times
  \int_0^{\pi} \frac{d\theta \sin\theta}{2}
  e^{-i q v_{\rm F} \tau \cos \theta}  \nonumber \\ &=&
  \ln \frac{1 + i (\Delta/\hbar + \omega_{\vq}^{\rm sw} - q v_{\rm F}) \tau_{\rm el}}{1 + i (\Delta/\hbar + \omega_{\vq}^{\rm sw} + q v_{\rm F}) \tau_{\rm el}}.
\end{eqnarray}%
For small $q$ and $\hbar \omega_{\vq}^{\rm sq} \ll \Delta$, 
this expression takes the form
\begin{equation}
  \frac{1}{\tau_\phi} = \int \frac{d\vq}{(2 \pi)^3} \frac{4 J^2T \overline{s}}{\hbar\omega_{\bf q}^\text{sw}}
 \frac{D q^2}{\Delta^2
  [1 + (\Delta \tau_{\rm el}/\hbar)^2]},
\end{equation}
where $D = v_{\rm F}^2 \tau_{\rm el}/3$ is the diffusion constant.

\section{Diagrammatic calculation of dephasing time}
\label{sec:4}

We now outline a full diagrammatic calculation of the spin wave contribution to the dephasing time. In the language of diagrammatic perturbation theory, the propagation of electrons in a disordered metal is described by means of the diffuson and Cooperon propagators. These describe the phase-coherent propagation of multiply scattered electrons traveling along identical paths or along identical but time-reversed paths, respectively. In the case of a ferromagnet, the internal magnetic field results in rapid dephasing of electrons traveling along time-reversed paths, and the Cooperon is strongly suppressed. For this reason, the calculation below focuses on the diffuson propagator.

The diffuson propagator relevant for the calculation of conductance fluctuations reads
\begin{eqnarray}
  \lefteqn{
  \mathcal{D}_{\sigma,\sigma'}(\vr'-\vr;\varepsilon,\omega)} \\ \nonumber &=&
  \langle
  \overline{G^{\rm R}_{\sigma}(\vr',\vr,\varepsilon+\hbar\omega/2)}\,
  \overline{G^{\rm A}_{\sigma'}(\vr,\vr',\varepsilon-\hbar\omega/2)}
  \rangle,
  \label{eq:diff}
\end{eqnarray}
where the bar $\overline{\cdots}$ denotes an average with respect to quantum-mechanical and thermal fluctuations of the $d$-band electron spin density, whereas the brackets $\langle \cdots \rangle$ denote a disorder average with respect to the impurity potential $V$. Further $G^{\rm R}$ and $G^{\rm A}$ are retarded and advanced Green functions for the conduction electrons, respectively. When calculating conductance fluctuations, the two Green functions in (\ref{eq:diff}) are part of interfering electronic trajectories at significantly different times. Since interactions with the spin density fluctuations along the two different trajectories are thus completely uncorrelated, the averaging over the fluctuations is done separately for the two Green functions.

We are interested in the Fourier transform
\begin{eqnarray}
  \mathcal{D}_{\sigma,\sigma'}(\vq;\varepsilon,\omega) &=&
  \int d\vrho\,
  \mathcal{D}_{\sigma,\sigma'}(\vrho;\varepsilon,\omega) 
  e^{i \vq \cdot \boldsymbol\rho}.
\end{eqnarray}
Since the dominant contribution to coherent propagation comes from constructively interfering trajectories of electrons with the same spin (due to the exchange splitting, electrons with different spin dephase rapidly), we will focus on the diagonal elements $\mathcal{D}_{\sigma,\sigma}(\vq;\varepsilon,\omega)$. For small $\vq$ and $\omega$ this diagonal propagator has the asymptotic parameter dependence
\begin{equation}
  \mathcal{D}_{\sigma,\sigma}(\vq;\varepsilon,\omega) \sim
  \frac{2\pi\nu}{\hbar}\frac{1}{D_{\sigma} q^2 - i \omega + 1/\tau_{\phi,\sigma}},
  \label{eq:asympt}
\end{equation}
where $D_\sigma = v_{\rm F}^2 \tau_{\sigma}/3$ is the diffusion constant. This parameter dependence corresponds to an exponential decay $\propto \exp(-\tau/\tau_{\phi,\sigma})$ of the Fourier transform of $\mathcal{D}_{\sigma,\sigma}(\vq;\varepsilon,\omega)$ at $\vq = 0$, similar to the exponential decay of the absolute value of the evolution matrix element $U_{\sigma\sigma}$ in the semiclassical picture of the previous Section.

\begin{figure}[t]
\begin{center}
\includegraphics[width=82mm]{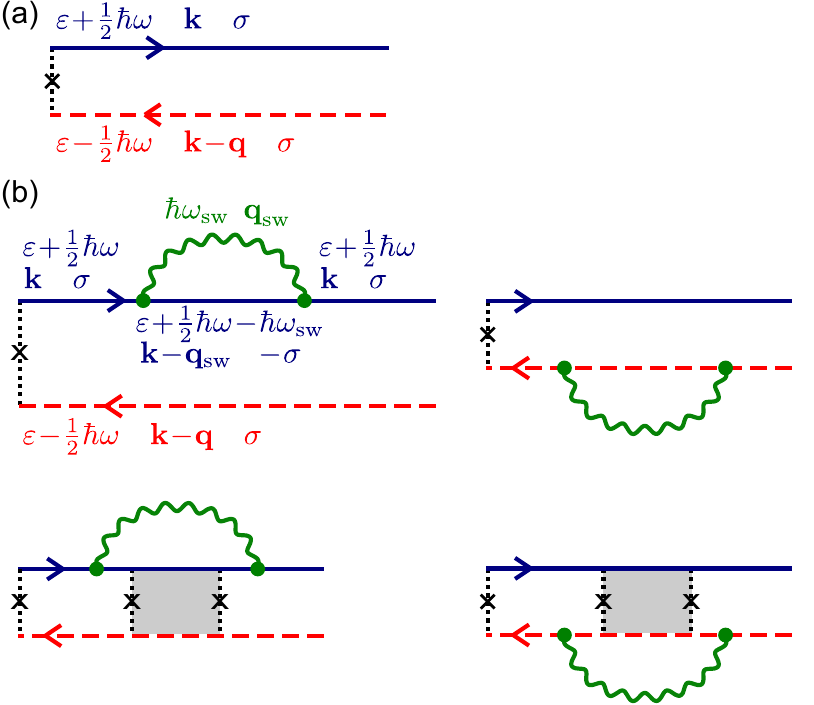}
\end{center}
\caption{(Color online) (a) Diagrammatic representation of the bare structure factor $\Pi^{(0)}_{\sigma,\sigma}$. The solid blue (dashed red) line represents a retarded (advanced) Green function, and the dotted line with the cross a correlated impurity scattering. (b) Diagrams representing the correction to the structure factor from spin wave induced dephasing. Correlated excitation and absorption of a spin wave changes temporarily the energy, momentum and spin of one of the propagators. The shaded block represents a ladder of impurity scatterings.}
\label{fig:diagrams}
\end{figure}

We denote the diffuson propagator in the absence of coupling to transverse fluctuations of the $d$ electron spin density by $\mathcal{D}_{\sigma,\sigma}^{(0)}(\vq;\varepsilon,\omega)$. This ``bare'' propagator is given by the equation
\begin{eqnarray}
  \mathcal{D}_{\sigma,\sigma}^{(0)}(\vq;\varepsilon,\omega) &=&
  \frac{2 \pi \nu \tau_{\sigma}}{\hbar} \frac{\Pi^{(0)}_{\sigma,\sigma}(\vq;\varepsilon,\omega)}{1 - \Pi^{(0)}_{\sigma,\sigma}(\vq;\varepsilon,\omega)},
\end{eqnarray}
where
\begin{eqnarray}
  \lefteqn{\Pi^{(0)}_{\sigma,\sigma}(\vq;\varepsilon,\omega)} 
  \nonumber \\ &=& 
  \frac{\hbar}{2\pi\nu \tau_{\sigma} V} 
  \\ && \mbox{} \times
  \sum_{\vk}
  \langle G^{\rm R}_\sigma (\vk,\varepsilon+\hbar\omega/2) \rangle
  \langle G^{\rm A}_\sigma (\vk-\vq,\varepsilon-\hbar\omega/2) \rangle
  \nonumber
\end{eqnarray}
is the bare structure factor, see Fig.\ \ref{fig:diagrams}a.
Substituting
\begin{eqnarray}
  \langle G^{\rm R}_\sigma (\vk,\varepsilon) \rangle &=&
  \frac{1}{\varepsilon - \varepsilon_{k} + \mu + \Delta \sigma/2 + \hbar i/2 \tau_{\sigma}}, \\
  \langle G^{\rm A}_\sigma (\vk,\varepsilon) \rangle &=&
  \frac{1}{\varepsilon - \varepsilon_{k} + \mu + \Delta \sigma/2 - \hbar i/2 \tau_{\sigma}},
\end{eqnarray}
for the impurity-averaged single-particle retarded and advanced Green functions, one finds the diagonal element
\begin{eqnarray}
  \Pi^{(0)}_{\sigma,\sigma}(\vq;\varepsilon,\omega) &=&
  \frac{1}{2 i q v_{\rm F} \tau_{\sigma}}
  \ln \frac{1 - i (\omega - q v_{\rm F})\tau_{\sigma}}
           {1 - i (\omega + q v_{\rm F})\tau_{\sigma}}
  \label{eq:Pi}
\end{eqnarray}
in the limit $q \ll k_{\rm F}$, $\hbar |\omega| \ll \mu$,
which leads to the asymptotic small-$q$ and small-$\omega$ dependence of Eq.\ (\ref{eq:asympt}) without the dephasing term.

Coupling to spin waves alters the structure factor, ultimately leading to a finite dephasing time in the diffuson propagator (\ref{eq:diff}). A diagrammatic evaluation, details of which can be found in the Appendix, leads to two types of corrections to the structure factor. The first correction is a spin-wave-mediated renormalization of the diffusion constant, which does not alter the functional form of the low-$q$ and low-$\omega$ asymptotics of the diffusion propagator. This correction is the equivalent of the Altshuler-Aronov correction to the conductivity from electron-electron interactions. Since this is a completely elastic correction, which does not contribute to dephasing, it will not be discussed here. The second correction is the spin-wave-mediated contribution to the dephasing rate. The relevant diagrams for this correction are shown in Fig.\ \ref{fig:diagrams}b. Denoting this contribution by $\Pi^{\rm sw}(\vq;\varepsilon,\omega)$ and analyzing the asymptotic dependence of the diffuson propagator for small $q$ and small $\omega$, one finds that
\begin{equation}
  \tau_{\phi,\sigma} = - \frac{\tau_{\sigma}}{\Pi_{\sigma,\sigma}^{\rm sw}(0;\varepsilon,0)}.
  \label{eq:dephup}
\end{equation}
Evaluation of the diagrams in Fig.\ \ref{fig:diagrams}b yields
\begin{eqnarray}
\lefteqn{\Pi^{{\rm sw}}_{\sigma,\sigma}(0;\varepsilon,0)} \nonumber \\ &=&
  \frac{2 J^2}{V}
  \sum_{\vq_{\rm sw}}
  \int \frac{d\omega_{\rm sw}}{2 \pi}
  \mbox{Im}\, \{\chi_{-\sigma,\sigma}^{\rm R}(\vq_{\rm sw},\omega_{\rm sw})\}
  \nonumber \\ && \mbox{} \times
  \left( \coth \frac{\hbar \omega_{\rm sw}}{2 T} + 
         \tanh \frac{\varepsilon - \hbar \omega_{\rm sw}}{2 T} \right)
  \nonumber \\ && \mbox{} \times
  \mbox{Re}\, \left\{ \vphantom{\frac{\Pi^{(0)}_0}{\Pi^{(0)}_0}}
  \Pi^{(0)}_{\sigma,-\sigma}(\vq_{\rm sw};\varepsilon-\hbar \omega_{\rm sw}/2,\omega_{\rm sw},2) 
  \right. \nonumber \\ && \left. \mbox{}
  + \frac{\tau_{\sigma} \Pi^{(0)}_{\sigma,-\sigma}(\vq_{\rm sw};\varepsilon-\hbar \omega_{\rm sw}/2,\omega_{\rm sw},1)^2}{\tau_{0}
  - \tau_{\sigma} \Pi^{(0)}_{\sigma,-\sigma}(\vq_{\rm sw};\varepsilon-\hbar\omega_{\rm sw}/2,\omega_{\rm sw},0)} \right\},  
  \label{eq:Piresult}
\end{eqnarray}
where we abbreviated
\begin{eqnarray}
  \lefteqn{\Pi^{(0)}_{\sigma,-\sigma}(\vq;\varepsilon,\omega;n)} \nonumber \\ &=&
  \frac{\hbar}{2 \pi \nu \tau_{\sigma} V}
  \sum_{\vk} 
    \langle G^{\rm R}_{\sigma}(\vk,\varepsilon + \hbar\omega/2) \rangle
  \nonumber \\ && \mbox{} \times
    \langle G^{\rm A}_{\sigma}(\vk,\varepsilon + \hbar\omega/2) \rangle^n
    \langle G^{\rm A}_{-\sigma}(\vk-\vq,\varepsilon-\hbar \omega/2) \rangle
  \nonumber \\ &=&
  \frac{(i\tau_{\sigma})^{n-1}}{2\hbar^n q v_{\rm F}}
  \ln \frac{1 - i (\omega + \Delta \sigma/\hbar - q v_{\rm F})\tau_{\rm a}}
           {1 - i (\omega + \Delta \sigma/\hbar + q v_{\rm F})\tau_{\rm a}},  
\end{eqnarray}
with
\begin{equation}
  \frac{1}{\tau_0} =
  \frac{1}{2 \tau_{\uparrow}} + \frac{1}{2 \tau_{\downarrow}}
  - \frac{1}{\tau_{\rm d}},\ \
  \frac{1}{\tau_{\rm a}} 
  = \frac{1}{2 \tau_{\uparrow}} + \frac{1}{2 \tau_{\downarrow}},
\end{equation}
the scattering rate for correlated scattering events of majority and minority electrons and the mean scattering rate, respectively. Setting $\varepsilon=0$, we thus find that the spin wave contribution to the dephasing rate of conduction electrons at the Fermi level is
\begin{eqnarray}
  \frac{1}{\tau_{\phi,\sigma}} &=&
  \frac{1}{V} \sum_\vq 
  \frac{4 J^2 \overline{s}}{\hbar^2 \sinh(\hbar \omega_{\vq}^{\rm sw}/T)} 
  \mbox{Re}\,
  \frac{\tau_0 i K}{2 q v_{\rm F} \tau_0 - i K},
  \nonumber \\
  \label{eq:dtres}
\end{eqnarray}
with
\begin{equation}
  K = \ln \frac{1 + i (\Delta/\hbar + \omega^{\rm sw}_{\vq} - q v_{\rm F})\tau_{\rm a}}{1 + i (\Delta/\hbar + \omega^{\rm sw}_{\vq} + q v_{\rm F}) \tau_{\rm a}}.
\end{equation} 
This result is identical to the result of the semiclassical calculation, up to the replacement of the summation over $\vq$ by an integral and the replacements $\sinh(\hbar \omega_{\vq}^{\rm sw}/T) \to \hbar \omega_{\vq}^{\rm sw}/T$ and $\tau_{0},\tau_{\rm a} \to \tau_{\rm el}$. We note that, in contrast to previous calculations of the spin wave contribution to the dephasing rate,\cite{JPSJ.72.1155,muttalibwoelfle} the expression derived above neither relies on the diffusion approximation $q l_{\rm el} \ll 1$, where $l_{\rm el}$ is the elastic mean free path of the conduction electrons, nor on the dirty-limit condition $\Delta \tau_{\rm el}/\hbar \ll 1$.
  
\section{Temperature dependence of dephasing time in a wire geometry}
\label{sec:5}

The only ingredient still missing is the spin wave dispersion relation $\omega^\text{sw}_{\bf q}$. As long as the typical wave length of the spin waves is large compared to the Fermi wave length, macroscopic spin wave theory can be employed to find $\omega^\text{sw}_{\bf q}$. For definiteness, we assume the anisotropy to be uniaxial and restrict ourselves to a  quasi-one-dimensional (wire) geometry. Choosing the $z$-axis to be parallel with the easy axis of the magnet, one then has\cite{kittel}
\begin{widetext}
\begin{eqnarray}
  (\hbar\omega^\text{sw}_{\bf q})^2 &=&
  \Big(\hbar D^\text{sw}q^2 + g\mu_{\rm B} 
  B 
  + \frac{2g\mu_{\rm B}K}{M_s}
  + 4\pi \gamma M_s\sin^2 \theta_{\bf q} \Big)
  \Big(\hbar D^\text{sw}q^2 + g\mu_{\rm B} B
  + \frac{2g\mu_{\rm B}K}{M_s}\Big).
\label{eq:disp}
\end{eqnarray}
\end{widetext}
Here, $D^\text{sw}$ is the spin wave stiffness (usually of the order $D^\text{sw} \sim \Delta/\hbar k_F^2$), $g\mu_{\rm B} B$ is the electronic Zeeman splitting due to an externally applied magnetic field along the $z$ axis, $M_s = g\mu_{\rm B} \overline{s}$ is the saturation magnetization, $\gamma=g\mu_{\rm B}\mu_0$ is the gyromagnetic ratio, $K$ is the energy density characterizing the magnetocrystalline anisotropy, $\mu_0$ is the permeability of free space, and $\theta_{\bf q}$ denotes the angle between ${\bf q}$ and the $z$-axis. In general, the demagnetizing field should be added to the applied field, corresponding to the replacement $B \to B - \mu_0 M_s \xi/4 \pi$ in the equation above, where $\xi$ is a dimensionless constant determined by the shape of the ferromagnet, such that $\xi$ is of order unity for a reasonably symmetric shape, and $\xi \to 0$ in the limit of a quasi-one-dimensional geometry (length much longer than width $W$). [Note that this dispersion relation differs from the one used in Ref.\ \onlinecite{muttalibwoelfle}, $\hbar\omega_{\bf q}^{\rm sw} = E_G + Aq^2$. This (more phenomenological) dispersion relation has only two ingredients, a spin wave gap $E_G$ and a stiffness $A$, covering only a subset of the regimes we investigate in this work.]

Below, we will neglect the contribution of the anisotropy to the spin wave energy, supporting this assumption with typical parameters for iron ($K \approx 5 \times 10^4$~J/m$^3$, $\gamma M_s\approx 0.25~$meV, see Ref.\ \onlinecite{mmm}), giving $K/2\pi\mu_0 M_s^2 \approx 2\cdot 10^{-3}$.
Then, for the case of no externally applied field, the dispersion relation (\ref{eq:disp}) simplifies to
\begin{equation}
\hbar\omega^\text{sw}_{\bf q} = \sqrt{ \big(\hbar D^\text{sw}q^2 + 4\pi \gamma M_s\sin^2 \theta_{\bf q} \big)\hbar D^\text{sw}q^2}.
\label{eq:dispsimp}
\end{equation}

Spin waves with an energy $\hbar\omega_{\bf q}^\text{sw} \gtrsim T$ do not significantly contribute to dephasing. From Eq.\ (\ref{eq:dispsimp}) we then conclude that the order of magnitude of the largest momenta which have to be taken into account in the calculation of the dephasing rate satisfies
\begin{equation}
  q_{\rm max} \sim \sqrt{T/\hbar D^\text{sw}}.
  \label{eq:qmax}
\end{equation}
The calculation of the previous section is thus valid as long as $q_{\rm max} \ll k_{\rm F}$, which is satisfied for all $T \ll \Delta$. A stronger constraint has to be met in order to justify the use of the diffusion approximation for the conduction electrons, as employed in Refs \onlinecite{JPSJ.72.1155} and \onlinecite{muttalibwoelfle}. This constraint is $q_{\rm max} \ll l_{\rm el}^{-1}$, which is met only if $T \ll \Delta (k_{\rm F} l_{\rm el})^{-2}$, imposing a severe restriction on the range of temperatures which can be investigated.

Eqs (\ref{eq:dtres})--(\ref{eq:disp}) form the most general result of our work: They allow to calculate (numerically) the dephasing time for arbitrary $\Delta$ (as long as $\Delta \gg T$) and, by including a demagnetizing field into (\ref{eq:disp}), for arbitrarily shaped samples. Let us now investigate different limits of $\Delta\tau_{\rm el}/\hbar$, in which the expression for the spin wave contribution to the dephasing rate simplifies considerably and we can arrive at explicit expressions for the dephasing rate.

\subsection{The limit $\Delta\tau_{\rm el}/\hbar \gg 1$, and $\tau_\ua \neq \tau_\da$}\label{subsec:dtlarge}

We first consider the most realistic case, i.e., where the exchange splitting $\Delta$ is large enough that $\Delta\tau_{\rm el}/\hbar \gg 1$ and where the elastic scattering times $\tau_{\uparrow}$ and $\tau_{\downarrow}$ are significantly different for the two spin directions. In this regime, we expand Eq.\ (\ref{eq:dtres}) to leading order in $\hbar/\Delta\tau_{\rm el}$,
\begin{equation}
  \frac{1}{\tau_{\phi,\sigma}} =
  \frac{1}{V} \sum_\vq
  \frac{4 J^2\overline{s}}{\Delta^2 \tau_{\rm d}
  \sinh(\hbar \omega^{\rm sw}_{\vq}/T)}.
\label{eq:dtres2}
\end{equation}
In the calculations that follow, we will replace the summation in Eq.\ (\ref{eq:dtres2}) by an integral and use the approximation (\ref{eq:dispsimp}) for the spin wave dispersion relation. This replacement and the use of this approximation requires a further discussion for two reasons. 
(i) The sum in Eq.\ (\ref{eq:dtres2}) diverges for small $q$ in a wire geometry if we use Eq.\ (\ref{eq:dispsimp}) for $\omega^{\rm sw}_{\vq}$. The origin of the divergence is that $\omega^{\rm sw}_{\vq} \to 0$ for $q \to 0$ at zero magnetic field when the anisotropy $K$ and demagnetizing factor $\xi$ are set to zero. The divergence is removed if we consider the full dispersion relation of Eq.\ (\ref{eq:disp}), for which $\omega^{\rm sw}_{\vq}$ takes the finite value $\omega_0^{\rm sw} \equiv \omega_K = 2 g \mu_{\rm B} K/\hbar M_s$ at $q = 0$ (and zero magnetic field). For sufficiently large anisotropy $\omega_K \gtrsim D^{\rm sw}/W^2$, the integrand has a smooth dependence on $\vq$ and the sum over $\vq$ may be replaced by an integral. The resulting integral, however, turns out to be largely independent of $K$ as long as $\hbar\omega_K \ll T$, which justifies the use of a spin wave dispersion relation with $K=0$ in combination with the replacement of the summation in Eq.\ (\ref{eq:dtres2}) by an integral in the parameter regime $\hbar D^{\rm sw}/W^2 \lesssim \hbar\omega_K \ll T$.
(ii) The replacement of the summation over $\vq$ by an integral is allowed only if  the transverse dimensions of the wire $W$ are small enough that $W \gg q^{-1}_{\rm max} \sim  k^{-1}_F\sqrt{\Delta/T}$. Taking realistic parameters for Fe and a wire width $W = 20$ nm, this limits the following calculations to temperatures $T \gg 0.08$ K.

We now focus on different limits where we can explicitly solve this integral. The result of the integral still depends on the ratio between the temperature and the demagnetization energy $4\pi \gamma M_s$. When we express the demagnetization energy in terms of temperature (e.g., $4\pi \gamma M_s \sim 3~$meV $\sim 35~$K for iron), we see that both the low-temperature limit $T \ll 4\pi \gamma M_s$ and the high-temperature limit $T \gg 4\pi \gamma M_s$ could be experimentally relevant.

We start by rewriting the integral version of (\ref{eq:dtres2}) as
\begin{equation}
  \frac{1}{\tau_{\phi,\sigma}} =
  A \int_0^\infty dx \int_0^{\pi /2} d\theta
  \frac{x^2 \sin \theta}{\sinh\Big(ax\sqrt{x^2 + \sin^2\theta}\Big)},
\label{eq:dtres2a}
\end{equation}
with $A = (2 J^2 \overline{s} /\pi^2 \Delta^2 \tau_{\rm d})(4\pi\gamma M_s / \hbar D^{\rm sw})^{3/2}$ and $a = 4\pi\gamma M_s/T$. In the low-temperature limit we have $a \gg 1$, and we can approximate $\sin \theta \approx \theta$ for all relevant $\theta$. We then can perform the integration, to arrive at
\begin{equation}
\frac{1}{\tau_{\phi,\sigma}} = C_1\frac{J^2T^{5/2}}{\Delta^2\tau_{\rm d}\gamma g\mu_{\rm B}(\hbar D^{\rm sw})^{3/2}}, \ \ 
  T \ll 4 \pi \gamma M_s,
\label{eq:dtres3}
\end{equation}
with $C_1 \approx 0.0473$ being a numerical prefactor.
The opposite case of large temperature has $a\ll 1$, and can be evaluated as well. In this case we can approximate $\sqrt{x^2 + \sin^2\theta} \approx x$ in the integrand of (\ref{eq:dtres2a}). The resulting integral can be solved, yielding
\begin{equation}
\frac{1}{\tau_{\phi,\sigma}} = C_2 \frac{J^2\overline{s}T^{3/2}}{\Delta^2\tau_{\rm d} (\hbar D^{\rm sw})^{3/2}},\ \
  T \gg 4 \pi \gamma M_s,
\label{eq:dtres4}
\end{equation}
where the numerical prefactor $C_2 \approx 0.303$.

Another limit which can be considered is that of a large external magnetic field (but still $\Delta \gg g \mu_{\rm B} B$). When $g\mu_{\rm B} B \gg 4\pi\gamma M_s$, we can approximate $\hbar\omega_{\bf q}^\text{sw} \approx \hbar D^\text{sw}q^2 + g\mu_{\rm B} B$, independent of $\theta_{\bf q}$. For high temperatures $T \gg g\mu_{\rm B} B$ we then recover the high-temperature result of Eq.\ (\ref{eq:dtres4}). In the low-temperature limit $T \ll g\mu_{\rm B} B$ all spin wave energies contributing to the integral in Eq.\ (\ref{eq:dtres}) are larger than $T$, so we approximate $1/\sinh (\hbar \omega_{\bf q}^{\rm sw}/T) \approx 2e^{-\hbar\omega_{\bf q}^{\rm sw}/T}$. The resulting integral can be calculated, yielding a dephasing rate that is exponentially suppressed with increasing magnetic field $B$,
\begin{equation}
\frac{1}{\tau_{\phi,\sigma}} = \frac{1}{\pi^{3/2}}\frac{J^2\overline{s}T^{3/2}}{\Delta^2\tau_{\rm d}(\hbar D^{\rm sw})^{3/2}} e^{-g \mu_{\rm B} B/T},\ \
  T \ll g \mu_{\rm B} B.
\label{eq:dtres5}
\end{equation}

\subsection{The limit $\Delta\tau_{\rm el}/\hbar \gg 1$, and $\tau_\ua = \tau_\da$}

For clean ferromagnets, for which $\Delta\tau_{\rm el}/\hbar$ is still large, but not large enough to cause the elastic scattering times for majority and minority electrons to be significantly different from each other, we can use $\tau_\ua = \tau_\da = \tau_0 = \tau_a \equiv \tau_{\rm el}$ and $\tau_d \to \infty$. We see that in this case the leading order contribution we found in Eq.\ (\ref{eq:dtres2}) vanishes and we have to expand Eq.\ (\ref{eq:dtres}) to higher orders in $\hbar/\Delta\tau_{\rm el}$. Again limiting our calculation to the regime $W \gg k_{\rm F}^{-1} \sqrt{\Delta/T}$, the leading-order contribution to the dephasing rate is found to be
\begin{equation}
  \frac{1}{\tau_{\phi,\sigma}} =
  \int \frac{d\vq}{(2 \pi)^3}
  \frac{4 \hbar^2 J^2 \overline{s}\tau_{\rm el}}{(\Delta \tau_{\rm el})^4}\frac{ D q^2 \tau_{\rm el}}{\sinh(\hbar \omega^{\rm sw}_{\vq}/T)},
\label{eq:redld}
\end{equation}
where $D = v_{\rm F}^2 \tau_{\rm el}/3$ is the conduction electron diffusion constant. Asymptotic expressions for the dephasing rate can then be obtained in the same limits as treated in the previous subsection. In the low-temperature limit $T \ll 4\pi\gamma M_s$, we find
\begin{equation}
\frac{1}{\tau_{\phi,\sigma}} = C_3\frac{\hbar^2 J^2 D T^{7/2}}{\Delta^4\tau_{\rm el}^2 \gamma g\mu_{\rm B}(\hbar D^{\rm sw})^{5/2}}, \ \ 
  T \ll 4 \pi \gamma M_s,
\label{eq:dtres6}
\end{equation}
with $C_3 \approx 0.0367$.
In the opposite case of large temperature $T\gg  4\pi \gamma M_s$ we find
\begin{equation}
\frac{1}{\tau_{\phi,\sigma}} = C_4 \frac{\hbar^2 J^2 D \overline{s}T^{5/2}}{\Delta^4\tau_{\rm el}^2(\hbar D^{\rm sw})^{5/2}},\ \
  T \gg 4 \pi \gamma M_s,
\label{eq:dtres7}
\end{equation}
where the numerical prefactor $C_4 \approx 0.297$.
Finally, with a large external magnetic field, $g\mu_BB \gg 4\pi\gamma M_s$, and low temperatures, $T \ll g \mu_{\rm B} B$, we arrive at
\begin{equation}
\frac{1}{\tau_{\phi,\sigma}} = \frac{3}{2\pi^{3/2}}\frac{\hbar^2 J^2 D \overline{s}T^{5/2}}{\Delta^4\tau_{\rm el}^2(\hbar D^{\rm sw})^{5/2}} e^{-g \mu_{\rm B} B/T},\ \
  T \ll g \mu_B B.
\label{eq:dtres8}
\end{equation}

We note that the three explicit results in this Section are all a factor $\sim (\hbar/\Delta\tau_{\rm el})(T/\Delta)(D/D^{\rm sw})$ smaller than the corresponding results in the previous Section. The main reason for the large enhancement of dephasing when $\tau_\ua \neq \tau_\da$ lies in the fact that in this case electrons with different spins effectively see a different impurity potential. Constructive interference of propagating electrons with different spin is then not only suppressed by their energy difference $\Delta$, but also by the fact that their actual trajectories quickly diverge.

\subsection{The limit $\Delta\tau_{\rm el}/\hbar \ll 1$}

Finally, we consider the limit $\Delta\tau_{\rm el}/\hbar \ll 1$ of a dirty ferromagnet. We also assume here that $\Delta$ is too small to cause a significant difference between $\tau_\ua$ and $\tau_\da$, so we again set $\tau_\ua = \tau_\da = \tau_0 = \tau_a \equiv \tau_{\rm el}$ and $\tau_d \to \infty$. If $W \gg \max [k^{-1}_F\sqrt{\Delta/T}, l_{\rm el}\sqrt{\hbar/\Delta\tau_{\rm el}}]$ the summation over $\vq$ can be replaced by an integral and we find
\begin{equation}
  \frac{1}{\tau_{\phi,\sigma}} =
  \int \frac{d \vq}{(2\pi)^3}
  \frac{4 J^2 \overline{s}\tau_{\rm el} D q^2 \tau_{\rm el}}{[(\hbar D q^2 \tau_{\rm el})^2 + (\Delta \tau_{\rm el})^2] \sinh(\hbar \omega^{\rm sw}_{\vq}/T)}.
\label{eq:redsd2}
\end{equation}
This integral is difficult to evaluate; The most accurate way to find a dephasing time in a specific material is to solve it numerically. We can however arrive at an order-of-magnitude estimate and find the power-law dependence on $T$ in different regimes. Since $\Delta\tau_{\rm el}/\hbar \ll 1$ and $T\ll \Delta$ is a requirement for the validity of spin wave dispersion relation (\ref{eq:disp}), we only treat the low-temperature limit $T \ll 4\pi\gamma M_s$. Performing an analysis similar to the one leading to Eq.\ (\ref{eq:dtres3}), we find
\begin{widetext}
\begin{equation}
\frac{1}{\tau_{\phi,\sigma}} = \frac{J^2D^{\rm sw}\Delta^{3/2}}{\gamma g \mu_{\rm B}(\hbar D)^{5/2}}
\frac{1}{c^{3/2}} \int _0^{\infty }\frac{dx}{2\pi^3} \frac{2 x^{3/2} {\rm arcoth}(e^x)+\sqrt{x}\big[{\rm Li}_2(e^{-x})-{\rm Li}_2(-e^{-x})\big]}{x^2+c^2},
\end{equation}
\end{widetext}
with $c = \Delta D^{\rm sw}/T D \sim (\Delta/T)(\Delta\tau_{\rm el}/\hbar)(1/k_{\rm F}l_{\rm el})^2$ and ${\rm Li}_2$ the dilogarithmic function. Evaluating the integral for $c \ll 1$ and $c \gg 1$ we find
\begin{eqnarray}
\frac{1}{\tau_{\phi,\sigma}} &=& C_5 \frac{J^2}{\hbar ^2\gamma  g \mu_{\rm B} D^{\rm{sw}}}\frac{T^2}{\sqrt{\Delta  \hbar  D}},\ \  c \ll 1, \label{eq:dtres9} \\
\frac{1}{\tau_{\phi,\sigma}} &=& C_6 \frac{J^2D}{\gamma  g \mu_{\rm B}}\frac{T^{7/2}}{\Delta ^2\left(\hbar  D^{\rm{sw}}\right)^{5/2}},\ \  c \gg 1,
\label{eq:dtres10}
\end{eqnarray}
with $C_5 \approx 0.0884$ and $C_6 \approx 0.0514$ numerical constants.

We note that the same limit $\Delta\tau_{\rm el}/\hbar \ll 1$ has been considered previously by Takane.\cite{JPSJ.72.1155} Our result (\ref{eq:redsd2}) indeed coincides with Eq.\ (38) from Ref.\ \onlinecite{JPSJ.72.1155} in the limit of $\Delta \gg \hbar\omega^{\rm sw}_{\vq}$ and $\tau_\ua = \tau_\da$ (up to a factor $1/\pi$, which Takane overlooked in his last step). Our estimates (\ref{eq:dtres9}) and (\ref{eq:dtres10}) differ qualitatively from those in Ref.\ \onlinecite{JPSJ.72.1155} because ours have been derived consistently in the limit where $\tau_\ua$ and $\tau_\da$ are set equal, whereas Ref.\ \onlinecite{JPSJ.72.1155} considers the case of different scattering times for majority electrons and minority electrons.

\section{Conclusion}
\label{sec:6}

We calculated the contribution of the interaction between $s$-band electrons and $d$-band spin waves in ferromagnets to the dephasing rate of coherent propagation of the electrons. 
Our work, as opposed to previous calculations,\cite{JPSJ.72.1155,muttalibwoelfle} neither relies on the diffusion approximation $q l_{\rm el} \ll 1$, nor on the dirty-limit condition $\Delta \tau_{\rm el}/\hbar \ll 1$. We found qualitatively different results depending on whether $\Delta\tau_{\rm el}/\hbar \gg 1$ or $\Delta\tau_{\rm el}/\hbar \ll 1$.

We investigated the explicit temperature dependence of the dephasing rate in quasi-one-dimensional systems. In the most realistic limit of $\Delta\tau_{\rm el}/\hbar \gg 1$ we find that the temperature-dependent dephasing rate obeys a power law: at low temperatures it is proportional to $T^{5/2}$, whereas at higher temperatures it becomes proportional to $T^{3/2}$. If $\Delta\tau_{\rm el}/\hbar$ is still large, but $\Delta$ is not large enough to cause a significant difference between the effective disorder potential seen by electrons with spin up and down, the dephasing rate becomes proportional to $T^{7/2}$ at low temperatures and to $T^{5/2}$ at high temperatures.
We also provide expressions for general $\Delta\tau_{\rm el}/\hbar$ and arbitrarily shaped samples. In this general case however, the summation over ${\bf q}$ can not be performed in closed form, and our general expression has to be evaluated numerically.

A relevant question to answer is how the phase relaxation time thus found compares to the phase relaxation time due to electron-electron interactions. For a quasi-one-dimensional geometry, the latter is of the order~\cite{aa:book}
\begin{equation}
\frac{1}{\tau_{\varphi,{\rm ee}}} \sim \left(\frac{T}{D^{1/2}\nu_1\hbar^2}\right)^{2/3} \sim \left(\frac{1}{k_F^2A}\frac{T}{\hbar\sqrt{\tau_{\rm el}}}\right)^{2/3},
\label{eq:dephee}
\end{equation}
where $\nu_1$ is the effective one-dimensional density of states and $A$ is the cross-sectional area of the wire. Since dephasing from electron-electron interactions scales $\propto T^{2/3}$, which is a lower power of $T$ than any of the low-temperature limits considered in Sec.\ \ref{sec:5}, Coulomb interactions provide the dominant source of dephasing in the limit of sufficiently low temperatures. On the other hand, for sufficiently high temperatures, dephasing from spin waves may take over. 

We now estimate the temperature $T_{\rm min}$ above which the spin-wave-induced dephasing dominates for a wire with a cross section $A = (20\, \mbox{nm})^2$, using realistic parameters for Fe. Hereto, we take $\Delta \sim 9\times 10^3$~K (using parameters from Ref.\ \onlinecite{PhysRevB.66.024433} and assuming that roughly $\Delta \sim J\overline{s}$) and $\tau_{\rm el} \sim 3 \times 10^{-14}$~s $\sim (250\ \mbox{K})^{-1}$, which corresponds to the clean limit $\Delta\tau_{\rm el}/\hbar \gg 1$, for which the results of Sec.\ \ref{subsec:dtlarge} apply. We use our low-temperature result (\ref{eq:dtres3}), estimate $D^{\rm sw} \sim \Delta/\hbar k_F^2$ and assume $\tau_{\rm d} \sim \tau_{\rm el}$, which, together with typical parameters for iron,\cite{ashmer,PhysRevB.66.024433} gives $T_{\rm min} \sim 3.5$~K.

We have to make several consistency checks here: (i) The temperature found is consistent with our choice for the low-temperature regime $T\ll 4\pi\gamma M_s$ required for the validity of Eq.\ (\ref{eq:dtres3}). (ii) The transverse dimensions of the wire should exceed $k_F^{-1}\sqrt{\Delta/T_{\rm min}} \sim 3$~nm, which is indeed the case. (iii) To see whether the phase relaxation time at this transition temperature is still compatible with the quasi-one-dimensional limit of Eq.\ (\ref{eq:dephee}), we evaluate the dephasing length $\sqrt{D\tau_{\varphi,{\rm ee}}}$ with the numbers used at temperature $T = 3.5~K$: We find $\sqrt{D\tau_{\varphi,{\rm ee}}} \sim 10~\mu$m, indeed much larger than the chosen wire thickness of $\sim 20~$nm. Although the precise value of $T_{\rm min}$ depends on effects not taken into account here, such as the detailed band structure of Fe, or specifics of the impurities, the order-of-magnitude estimate $T_{\rm min} \sim 3.5$~K leaves room for a significant parameter regime where the interaction with spin waves could be the dominant mechanism of electronic dephasing.

Comparing our results with the available experiments,\cite{kn:kasai2003,kn:lee2007,kn:wei2006,kn:vila2007,kn:lee2004,kn:bolotin2006,kn:neumaier2008} we make the following observations: (i) The anomalously short dephasing lengths observed in some of the experiments (Refs \onlinecite{kn:kasai2003,kn:lee2004,kn:neumaier2008} report $L_\phi\sim$ 100--300 nm at 30--80 mK, and $L_{\phi} \sim 10$ nm at temperatures around 1 K) differ from the estimated rate for spin-wave-induced dephasing in Fe by three orders of magnitude. It seems unlikely that this large difference can be attributed to material-specific details not taken into account in the theory. Hence, we believe that spin waves are not the source of the anomalously high dephasing rates seen in these experiments. (ii) The temperature dependence of $1/\tau_\phi$ from experiments on quasi-one-dimensional ferromagnetic metal wires\cite{kn:lee2007,kn:neumaier2008} could fit a $T^{3/2}$-dependence in some data sets. A problem with extracting this power law dependence from the data is that the value of the dephasing length is so short (see above), that $L_\phi$ becomes comparable to the transverse dimensions of the wires in a significant range of temperatures. This makes it difficult to convert the measured quantity (the rms value of the conductance fluctuations or the correlation field or bias voltage) unambiguously into a temperature dependence of $L_\phi$, since the relation between these quantities and $L_\phi$ depends qualitatively on the effective dimensionality of the system.\cite{PhysRevLett.62.195,PhysRevB.35.1039} (iii) The measured indicators of quantum coherence (such as noise power of time-dependent conductance fluctuations, \cite{kn:lee2004} or correlation bias voltage\cite{kn:neumaier2008}) seem to be largely invariant when the external magnetic field is increased over a range of several Tesla. This, too, seems to indicate that the interaction with spin waves is not the dominant mechanism of dephasing at low temperatures, since one expects an exponential suppression of the spin-wave-induced dephasing rate when $g\mu_BB \gtrsim T$. However, at temperatures $T > T_{\rm min}$, where we theoretically expect spin waves to start playing a role (we estimated $T_{\rm min} \sim 3.5$~K, see above, which corresponds to magnetic fields $\sim 2.6$~T), one indeed would need a field of several Tesla before any field-induced suppression of the effect would become noticeable.

We close with a remark comparing the present calculation and the calculation of the dephasing time from electron-electron interactions.\cite{aa:book} There are two important differences between these two calculations. First, for Coulomb interactions the dephasing rate for a quasi-one-dimensional geometry can not be obtained in perturbation theory, but instead needs a self-consistency argument to cut off an otherwise divergent integral. No such divergence occurred in the present calculation. Second, the calculation of the dephasing rate from electron-electron interactions that enters the conductance fluctuations requires the calculation of a more complicated object than the simple diffuson propagator we consider.\cite{PhysRevB.65.115317} The origin of the difference is twofold: (i) The Coulomb interaction becomes effectively long range at low frequencies $\omega$, whereas the spin-wave-mediated interaction remains short range even at small $\omega$ because of the exchange splitting $\Delta$. Indeed, since every interaction with the spin waves involves an electronic spin flip, correlations over distances larger than $\min[\hbar v_{\rm F}/\Delta,(\hbar D/\Delta)^{1/2}]$ are strongly suppressed by the exchange splitting. (ii) The divergence at $\vq = 0$ and the resulting $ \omega_0^{\rm sw} = 0$ in our calculation (corresponding to a static magnetization pointing in a random direction) is cut off by the magnetocrystalline anisotropy, physically pinning the magnetization to a fixed direction.

\acknowledgments

We gratefully acknowledge discussions with Peter Silvestrov. This work is supported by the Alexander von Humboldt Foundation in the framework of the Alexander von Humboldt Professorship programme, endowed by the Federal Ministry of Education and Research.

\appendix 

\section{Evaluation of the diagrams}

\begin{figure}[!t]
\begin{center}
\includegraphics[width=82mm]{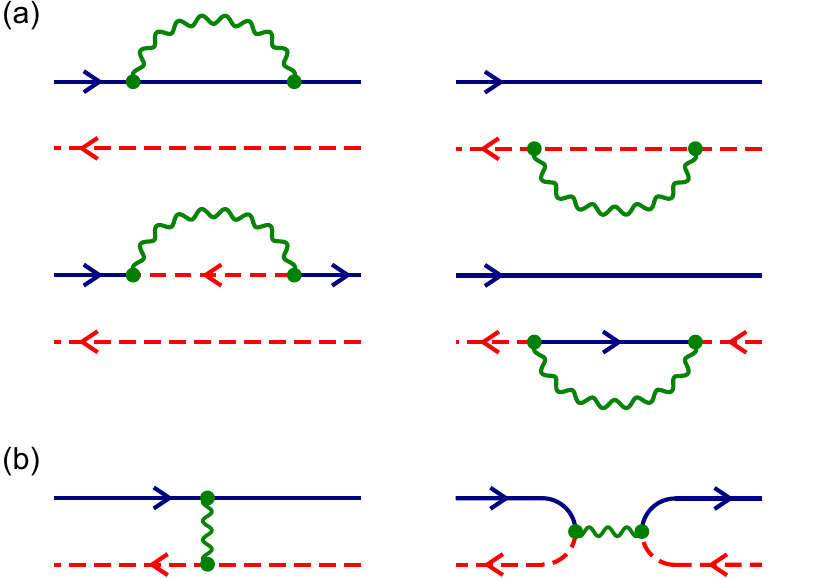}
\caption{(Color online) Diagrams representing the spin wave correction to the diffusion propagator. (Impurity lines are not shown in the diagrams in this figure.) (a) Diagrams in which the spin wave propagator does not connect the retarded and advanced propagators in the diffuson. (b) Diagrams in which retarded and advanced propagators are connected by a spin wave propagator. Only diagrams shown in part (a) are relevant for the calculation of the dephasing rate.}
\label{fig:app}
\end{center}
\end{figure}

The diffuson propagator describes the coherent propagation of two quantum mechanical probability amplitudes. The effect of spin waves on the propagation is qualitatively different for the cases that these amplitudes describe propagation through the sample at the same time, or at different times. The former case is relevant, e.g., for a calculation of the disorder-averaged conductance, whereas the latter scenario is relevant, e.g., for the calculation of the conductance fluctuations.\cite{PhysRevB.65.115317} Diagrammatically, the difference between the two cases is that all spin-wave-mediated interaction lines have to be included in the former case (diagrams of Fig.\ \ref{fig:app}a and b), whereas in the latter case only interaction lines that connect the retarded propagator to itself or that connect the advanced propagator to itself have to be considered (diagrams of Fig.\ \ref{fig:app}a). Impurity lines, on the other hand, can connect all Green functions. Our calculation of the dephasing rate applies to the second scenario. This is consistent with the definition of the diffuson propagator in Eq.\ (\ref{eq:diff}), where the spin wave averages are taken for each Green function separately and the disorder average is taken of the product of the two Green functions. (We note that current conservation protects the pole of the diffusion propagator in the former case.)

For our calculation of the diffuson propagator, we first consider the most general quantity
\begin{eqnarray}
  \lefteqn{
  {\cal D}_{\sigma_3,\sigma_4}^{\sigma_1,\sigma_2}(\vr'-\vr;i \varepsilon_n,i \varepsilon_m)}
  \nonumber \\ &=&
  \langle \overline{{\cal G}_{\sigma_1,\sigma_2}(\vr',\vr,i \varepsilon_n)}\, \overline{{\cal G}_{\sigma_3,\sigma_4}(\vr',\vr,i \varepsilon_m)} \rangle,
\end{eqnarray}
where $\varepsilon_n > 0$ and $\varepsilon_m < 0$ are fermionic Matsubara frequencies. Because the spin wave average is taken for each Green function separately, the Green functions themselves contain no off-diagonal elements, i.e., $\sigma_1 = \sigma_2$ and $\sigma_3=\sigma_4$, so we write the diffuson as
\begin{eqnarray}
  \lefteqn{
  {\cal D}_{\sigma,\sigma'}(\vr'-\vr;i \varepsilon_n,i \varepsilon_m)}
  \nonumber \\ &=&
  \langle \overline{{\cal G}_{\sigma}(\vr',\vr,i \varepsilon_n)}\, \overline{{\cal G}_{\sigma'}(\vr',\vr,i \varepsilon_m)} \rangle.
\end{eqnarray}
We further focus on the dominating contribution where $\sigma = \sigma'$, i.e., the diagonal elements of the diffusion propagator. These are then found to obey the Dyson equation
\begin{eqnarray}
  {\cal D}_{\sigma,\sigma}(\vr'-\vr;i \varepsilon_n,i \varepsilon_m) &=&
  2 \pi \nu \tau_{\sigma}
  \Pi_{\sigma,\sigma}(\vr'-\vr;i \varepsilon_n,i \varepsilon_m) 
  \nonumber \\ && \mbox{} +
  \int d\vr''
  \Pi_{\sigma,\sigma}(\vr''-\vr;i \varepsilon_n,i \varepsilon_m)
  \nonumber \\ && \ \ \mbox{} \times
  {\cal D}_{\sigma,\sigma}(\vr'-\vr'';i \varepsilon_n,i \varepsilon_m),
  \nonumber \\
  \label{eq:DPi}
\end{eqnarray}
where $\Pi_{\sigma,\sigma}(\vr'-\vr;i \varepsilon_n,i \varepsilon_m)$ is the full structure factor. Fourier transforming Eq.\ (\ref{eq:DPi}), we find
\begin{eqnarray}
  {\cal D}_{\sigma,\sigma}(\vq;i \varepsilon_n,i \varepsilon_m) &=&
  2 \pi \nu \tau_{\sigma}
  \Pi_{\sigma,\sigma}(\vq;i \varepsilon_n,i \varepsilon_m) 
   \\ && \mbox{} \times
  \left[1 + (2 \pi \nu \tau_{\sigma})^{-1} 
  {\cal D}_{\sigma,\sigma}(\vq;i \varepsilon_n,i \varepsilon_m)
  \right]. \nonumber
\end{eqnarray}  
For the calculation of the dephasing time, we need the simultaneous limits $\vq \to 0$, $\varepsilon_n \to \varepsilon + i 0$, and $\varepsilon_m \to \varepsilon - i 0$.

The diagrams contributing to $\Pi^{\rm sw}_{\sigma,\sigma}(\vq;i \varepsilon_n,i \varepsilon_m)$ are shown in Fig.\ \ref{fig:app2}. Evaluating these diagrams yields
\begin{widetext}
\begin{eqnarray}
&& \mbox{} \hspace{-.5cm}\Pi^{\rm sw}_{\sigma,\sigma}(\vq;i \varepsilon_n,i \varepsilon_m) =
\nonumber \\ && 
\frac{J^2T}{\hbar} \sum_{\nu_s<\varepsilon_n} \frac{1}{V}\sum_{{\bf q}_{\rm sw}} \chi_{-+}({\bf q}_{\rm sw},i\nu_s)\bigg\{ \frac{\hbar}{2\pi\nu\tau_\sigma V}\sum_{{\bf k}}{\cal G}_\sigma ({\bf k},i\varepsilon_n)^2{\cal G}_\sigma ({\bf k}-{\bf q},i\varepsilon_m) {\cal G}_{-\sigma} ({\bf k}-{\bf q}_{\rm sw},i\varepsilon_n-i\nu_s)
\nonumber \\ &&
\hspace{5cm}+\frac{\tau_\sigma \big[\frac{\hbar}{2\pi\nu\tau_\sigma V}\sum_{{\bf k}} {\cal G}_\sigma ({\bf k},i\varepsilon_n) {\cal G}_{-\sigma} ({\bf k}-{\bf q}_{\rm sw},i\varepsilon_n-i\nu_s){\cal G}_\sigma ({\bf k}-{\bf q},i\varepsilon_m)\big]^2}{\tau_0 - \tau_\sigma\frac{\hbar}{2\pi\nu\tau_\sigma V}\sum_{{\bf k}} {\cal G}_{-\sigma} ({\bf k}-{\bf q}_{\rm sw},i\varepsilon_n-i\nu_s){\cal G}_\sigma ({\bf k}-{\bf q},i\varepsilon_m)}\bigg\}
\nonumber \\ && 
+ \frac{J^2T}{\hbar} \sum_{\nu_s>\varepsilon_m} \frac{1}{V}\sum_{{\bf q}_{\rm sw}}\chi_{-+}({\bf q}_{\rm sw},i\nu_s)\bigg\{ \frac{\hbar}{2\pi\nu\tau_\sigma V}\sum_{{\bf k}} {\cal G}_\sigma ({\bf k},i\varepsilon_n) {\cal G}_\sigma ({\bf k}-{\bf q},i\varepsilon_m)^2{\cal G}_{-\sigma} ({\bf k}-{\bf q}-{\bf q}_{\rm sw},i\varepsilon_m-i\nu_s)
\label{eq:matsfull} \\ && 
\hspace{5.4cm}+\frac{\tau_\sigma \big[\frac{\hbar}{2\pi\nu\tau_\sigma V}\sum_{{\bf k}} {\cal G}_\sigma ({\bf k},i\varepsilon_n) {\cal G}_{-\sigma} ({\bf k}-{\bf q}-{\bf q}_{\rm sw},i\varepsilon_m-i\nu_s){\cal G}_\sigma ({\bf k}-{\bf q},i\varepsilon_m)\big]^2}{\tau_0 - \tau_\sigma\frac{\hbar}{2\pi\nu\tau_\sigma V}\sum_{{\bf k}} {\cal G}_\sigma ({\bf k},i\varepsilon_n) {\cal G}_{-\sigma} ({\bf k}-{\bf q}-{\bf q}_{\rm sw},i\varepsilon_m-i\nu_s)} \bigg\}
\nonumber \\ && 
+ \frac{J^2T}{\hbar} \sum_{\nu_s<\varepsilon_m} \frac{1}{V}\sum_{{\bf q}_{\rm sw}} \chi_{-+}({\bf q}_{\rm sw},i\nu_s) {\cal H}_\sigma ({\bf q}_{\rm sw},-{\bf q};i\varepsilon_m,i\varepsilon_n)
\Big[\frac{\tau_0}{\tau_0-\tau_\sigma\frac{\hbar}{2\pi\nu\tau_\sigma V}\sum_{\bf k} {\cal G}_\sigma ({\bf k},i\varepsilon_m) {\cal G}_{-\sigma} ({\bf k}-{\bf q}_{\rm sw},i\varepsilon_m-i\nu_s)} \Big]^2
\nonumber \\ && 
+ \frac{J^2T}{\hbar} \sum_{\nu_s>\varepsilon_n} \frac{1}{V}\sum_{{\bf q}_{\rm sw}} \chi_{-+}({\bf q}_{\rm sw},i\nu_s) {\cal H}_\sigma ({\bf q}_{\rm sw},{\bf q};i\varepsilon_n,i\varepsilon_m)
\Big[\frac{\tau_0}{\tau_0-\tau_\sigma\frac{\hbar}{2\pi\nu\tau_\sigma V}\sum_{\bf k} {\cal G}_\sigma ({\bf k},i\varepsilon_n) {\cal G}_{-\sigma} ({\bf k}-{\bf q}_{\rm sw},i\varepsilon_n-i\nu_s)}\Big]^2. \nonumber
\end{eqnarray}
\begin{figure*}[t]
\includegraphics[width=180mm]{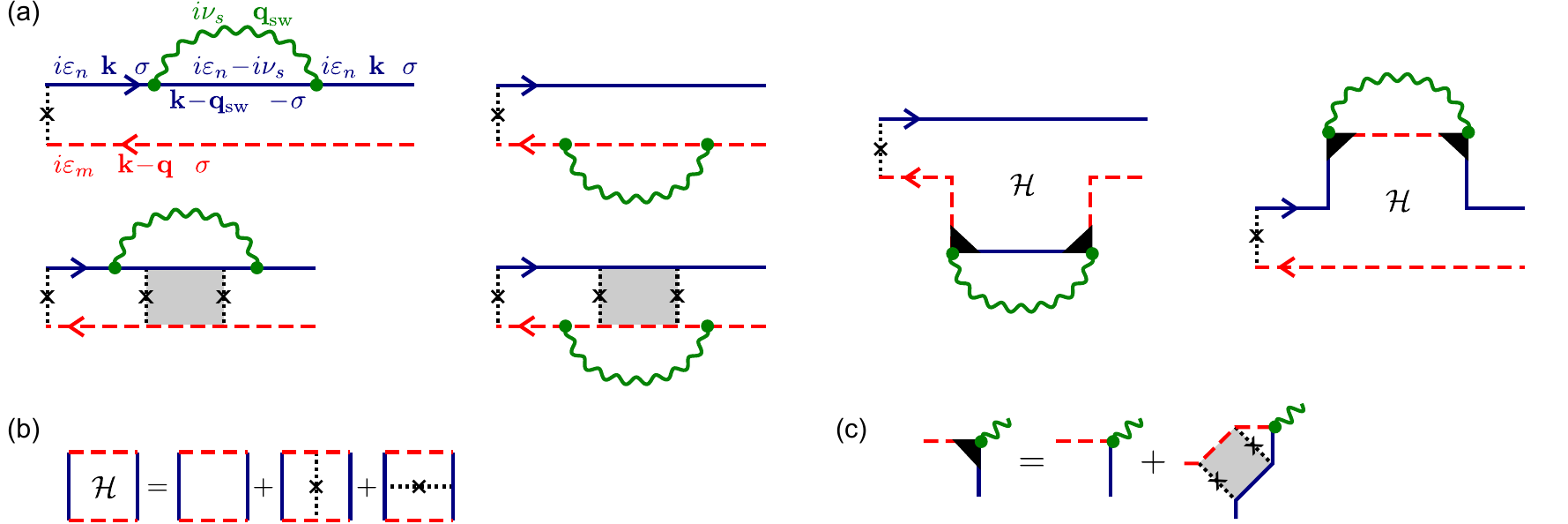}
\begin{center}
\caption{(Color online) (a) Diagrams representing the spin wave correction to the disorder-averaged structure factor. (b) Diagrammatic definition of the Hikami box appearing in some of the diagrams of part (a). (c) Definition of the dressed vertices appearing in some of the diagrams of part (a).}
\label{fig:app2}
\end{center}
\end{figure*}%
\noindent The impurity averaged Green functions $\mathcal{G}_\sigma({\bf k},i\varepsilon)$ and response function $\chi_{-+}({\bf q},i\nu_s)$ are all written in Matsubara representation.
The contribution to (\ref{eq:matsfull}) by the Hikami boxes (see Fig.\ \ref{fig:app}b) is
\begin{eqnarray}
\mbox{}{\cal H}_\sigma ({\bf q}_{\rm sw},{\bf q};i\varepsilon_n,i\varepsilon_m)  &=&
\frac{\hbar}{2\pi\nu\tau_\sigma V}\sum_{{\bf k}}  {\cal G}_\sigma ({\bf k},i\varepsilon_n)^2{\cal G}_\sigma ({\bf k}-{\bf q},i\varepsilon_m) {\cal G}_{-\sigma} ({\bf k}-{\bf q}_{\rm sw},i\varepsilon_n-i\nu_s)
\nonumber \\ && \quad
+ \frac{\tau_\sigma}{\tau_0}\Big[ \frac{\hbar}{2\pi\nu\tau_\sigma V} \sum_{{\bf k}}  {\cal G}_\sigma ({\bf k},i\varepsilon_n){\cal G}_\sigma ({\bf k}-{\bf q},i\varepsilon_m) {\cal G}_{-\sigma} ({\bf k}-{\bf q}_{\rm sw},i\varepsilon_n-i\nu_s)\Big]^2
\\ && \quad
+ \Big( \frac{\hbar}{2\pi\nu\tau_\sigma V}\Big)^2 \sum_{{\bf k},{\bf k}'}
{\cal G}_\sigma ({\bf k},i\varepsilon_n)^2{\cal G}_{-\sigma} ({\bf k}-{\bf q}_{\rm sw},i\varepsilon_n-i\nu_s)
{\cal G}_\sigma ({\bf k}',i\varepsilon_n)^2{\cal G}_\sigma ({\bf k}'-{\bf q},i\varepsilon_m). \nonumber 
\end{eqnarray}

We replace the summation over the intermediate Matsubara frequency $\nu_s$ by an integration over real frequencies and perform the analytical continuations  $i\varepsilon_n \to \varepsilon + \hbar \omega/2 + i 0$ and $i\varepsilon_m \to \varepsilon - \hbar \omega/2 - i 0$.
For the spin wave contribution to the structure factor we take the limit $\vq, \omega \to 0$ and find the expression
\begin{eqnarray}
\Pi^{{\rm sw}}_{\sigma,\sigma}(0;\varepsilon,0) &=&
  \frac{2 J^2}{V}
  \sum_{\vq_{\rm sw}}
  \int \frac{d\omega_{\rm sw}}{2 \pi}
  \left( \coth \frac{\hbar \omega_{\rm sw}}{2 T} + 
         \tanh \frac{\varepsilon - \hbar \omega_{\rm sw}}{2 T} \right)
  \mbox{Im}\, \{\chi_{-\sigma,\sigma}^{\rm R}(\vq_{\rm sw},\omega_{\rm sw})\}
\nonumber \\ && \quad \times
  \mbox{Re}\, \left\{ \vphantom{\frac{\Pi^{(0)}_0}{\Pi^{(0)}_0}}
  \Pi^{(0)}_{\sigma,-\sigma}(\vq_{\rm sw};\varepsilon-\hbar \omega_{\rm sw}/2,\omega_{\rm sw},2) 
  + \frac{\tau_{\sigma} \Pi^{(0)}_{\sigma,-\sigma}(\vq_{\rm sw};\varepsilon-\hbar \omega_{\rm sw}/2,\omega_{\rm sw},1)^2}{\tau_{0}
  - \tau_{\sigma} \Pi^{(0)}_{\sigma,-\sigma}(\vq_{\rm sw};\varepsilon-\hbar\omega_{\rm sw}/2,\omega_{\rm sw},0)} \right\} 
\nonumber \\ && -
  \frac{J^2}{V}
  \sum_{\vq_{\rm sw}}
  \int \frac{d\omega_{\rm sw}}{2 \pi}
  \tanh \frac{\varepsilon - \hbar \omega_{\rm sw}}{2 T}
\nonumber \\ && \quad \times
  \mbox{Im}\, \Bigg\{\chi_{-\sigma,\sigma}^{\rm R}(\vq_{\rm sw},\omega_{\rm sw})
  \bigg[\vphantom{\frac{\Pi^{(0)}_0}{\Pi^{(0)}_0}}
  \Pi^{(0)}_{\sigma,-\sigma}(\vq_{\rm sw};\varepsilon-\hbar \omega_{\rm sw}/2,\omega_{\rm sw},2) 
  + \frac{\tau_{\sigma} \Pi^{(0)}_{\sigma,-\sigma}(\vq_{\rm sw};\varepsilon-\hbar \omega_{\rm sw}/2,\omega_{\rm sw},1)^2}{\tau_{0}
  - \tau_{\sigma} \Pi^{(0)}_{\sigma,-\sigma}(\vq_{\rm sw};\varepsilon-\hbar\omega_{\rm sw}/2,\omega_{\rm sw},0)}
\nonumber \\ && \hspace{4cm}
+ H_{\sigma}({\bf q}_{\rm sw};\varepsilon,\omega_{\rm sw}) \frac{\tau_{0} }{\tau_{0}
  - \tau_{\sigma} \Pi^{(0)}_{\sigma,-\sigma}(\vq_{\rm sw};\varepsilon-\hbar\omega_{\rm sw}/2,\omega_{\rm sw},0)}\bigg]\Bigg\},
  \label{eq:Pifinal}
\end{eqnarray}
where we abbreviated as in the main text
\begin{eqnarray}
\Pi^{(0)}_{\sigma,-\sigma}(\vq;\varepsilon,\omega;n) = \frac{\hbar}{2 \pi \nu \tau_{\sigma} V}
  \sum_{\vk} 
    \langle G^{\rm R}_{\sigma}(\vk,\varepsilon + \hbar\omega/2) \rangle
    \langle G^{\rm A}_{\sigma}(\vk,\varepsilon + \hbar\omega/2) \rangle^n
    \langle G^{\rm A}_{-\sigma}(\vk-\vq,\varepsilon-\hbar \omega/2) \rangle,
\end{eqnarray}
and have written the contribution by the Hikami box as
\begin{eqnarray}
\mbox{}H_{\sigma}({\bf q}_{\rm sw};\varepsilon,\omega_{\rm sw}) &=&
\frac{\hbar}{2\pi\nu\tau_\sigma V}\sum_{{\bf k}}  \langle G^{\rm R}_\sigma ({\bf k},\varepsilon)\rangle^2 \langle G^{\rm A}_\sigma ({\bf k}-{\bf q},\varepsilon)\rangle \langle G^{\rm A}_{-\sigma} ({\bf k}-{\bf q}_{\rm sw},\varepsilon-\hbar\omega_{\rm sw})\rangle
\nonumber \\ && \quad
+ \frac{\tau_\sigma}{\tau_0}\Big[ \frac{\hbar}{2\pi\nu\tau_\sigma V} \sum_{{\bf k}} \langle G^{\rm R}_\sigma ({\bf k},\varepsilon)\rangle\langle G^{\rm A}_\sigma ({\bf k}-{\bf q},\varepsilon)\rangle\langle G^{\rm A}_{-\sigma} ({\bf k}-{\bf q}_{\rm sw},\varepsilon-\hbar\omega_{\rm sw})\rangle \Big]^2
\\ && \quad
+ \Big( \frac{\hbar}{2\pi\nu\tau_\sigma V}\Big)^2 \sum_{{\bf k},{\bf k}'}
\langle G^{\rm R}_\sigma ({\bf k},\varepsilon)\rangle^2 \langle G^{\rm A}_{-\sigma} ({\bf k}-{\bf q}_{\rm sw},\varepsilon-\hbar\omega_{\rm sw})\rangle
\langle G^{\rm R}_\sigma ({\bf k}',\varepsilon)\rangle^2 \langle G^{\rm A}_\sigma ({\bf k}'-{\bf q},\varepsilon)\rangle. \nonumber 
\end{eqnarray}
\end{widetext}
The first term of (\ref{eq:Pifinal}) is identical to Eq.\ (\ref{eq:Piresult}) of the main text; The second term vanishes upon performing the integration over $\vk$. (It represents a spin-wave-induced renormalization of the diffusion constant similar to the Altshuler-Aronov correction the the conductivity of a disordered metal, which is an elastic correction to the propagation of conduction electrons and does not cause dephasing.)


\begin{thebibliography}{36}
\expandafter\ifx\csname natexlab\endcsname\relax\def\natexlab#1{#1}\fi
\expandafter\ifx\csname bibnamefont\endcsname\relax
  \def\bibnamefont#1{#1}\fi
\expandafter\ifx\csname bibfnamefont\endcsname\relax
  \def\bibfnamefont#1{#1}\fi
\expandafter\ifx\csname citenamefont\endcsname\relax
  \def\citenamefont#1{#1}\fi
\expandafter\ifx\csname url\endcsname\relax
  \def\url#1{\texttt{#1}}\fi
\expandafter\ifx\csname urlprefix\endcsname\relax\def\urlprefix{URL }\fi
\providecommand{\bibinfo}[2]{#2}
\providecommand{\eprint}[2][]{\url{#2}}

\bibitem[{\citenamefont{Imry}(1997)}]{imry:book}
\bibinfo{author}{\bibfnamefont{Y.}~\bibnamefont{Imry}},
  \emph{\bibinfo{title}{Introduction to mesoscopic physics}}
  (\bibinfo{publisher}{Oxford University Press}, \bibinfo{year}{1997}).

\bibitem[{\citenamefont{Akkermans and Montambaux}(2007)}]{akkermans}
\bibinfo{author}{\bibfnamefont{E.}~\bibnamefont{Akkermans}} \bibnamefont{and}
  \bibinfo{author}{\bibfnamefont{G.}~\bibnamefont{Montambaux}},
  \emph{\bibinfo{title}{Mesoscopic Physics of Electrons and Photons}}
  (\bibinfo{publisher}{Cambridge University Press, Cambridge, UK},
  \bibinfo{year}{2007}).

\bibitem[{\citenamefont{Bergmann}(1984)}]{bergmann19841}
\bibinfo{author}{\bibfnamefont{G.}~\bibnamefont{Bergmann}},
  \bibinfo{journal}{Physics Reports} \textbf{\bibinfo{volume}{107}},
  \bibinfo{pages}{1 } (\bibinfo{year}{1984}).

\bibitem[{\citenamefont{Altshuler and Aronov}(1985)}]{aa:book}
\bibinfo{author}{\bibfnamefont{B.~L.} \bibnamefont{Altshuler}}
  \bibnamefont{and} \bibinfo{author}{\bibfnamefont{A.~G.}
  \bibnamefont{Aronov}}, in \emph{\bibinfo{booktitle}{Electron-electron
  interaction in disordered systems}} (\bibinfo{publisher}{North-Holland,
  Amsterdam}, \bibinfo{year}{1985}).

\bibitem[{\citenamefont{Lee and Ramakrishnan}(1985)}]{RevModPhys.57.287}
\bibinfo{author}{\bibfnamefont{P.~A.} \bibnamefont{Lee}} \bibnamefont{and}
  \bibinfo{author}{\bibfnamefont{T.~V.} \bibnamefont{Ramakrishnan}},
  \bibinfo{journal}{Rev. Mod. Phys.} \textbf{\bibinfo{volume}{57}},
  \bibinfo{pages}{287} (\bibinfo{year}{1985}).

\bibitem[{\citenamefont{B\"{u}ttiker et~al.}(1983)\citenamefont{B\"{u}ttiker,
  Imry, and Landauer}}]{buttiker:physlett}
\bibinfo{author}{\bibfnamefont{M.}~\bibnamefont{B\"{u}ttiker}},
  \bibinfo{author}{\bibfnamefont{Y.}~\bibnamefont{Imry}}, \bibnamefont{and}
  \bibinfo{author}{\bibfnamefont{R.}~\bibnamefont{Landauer}},
  \bibinfo{journal}{Phys. Lett.} \textbf{\bibinfo{volume}{96A}},
  \bibinfo{pages}{365} (\bibinfo{year}{1983}).

\bibitem[{\citenamefont{Webb et~al.}(1985)\citenamefont{Webb, Washburn, Umbach,
  and Laibowitz}}]{intel}
\bibinfo{author}{\bibfnamefont{R.~A.} \bibnamefont{Webb}},
  \bibinfo{author}{\bibfnamefont{S.}~\bibnamefont{Washburn}},
  \bibinfo{author}{\bibfnamefont{C.~P.} \bibnamefont{Umbach}},
  \bibnamefont{and} \bibinfo{author}{\bibfnamefont{R.~B.}
  \bibnamefont{Laibowitz}}, \bibinfo{journal}{Phys. Rev. Lett.}
  \textbf{\bibinfo{volume}{54}}, \bibinfo{pages}{2696} (\bibinfo{year}{1985}).

\bibitem[{\citenamefont{Lee and Stone}(1985)}]{PhysRevLett.55.1622}
\bibinfo{author}{\bibfnamefont{P.~A.} \bibnamefont{Lee}} \bibnamefont{and}
  \bibinfo{author}{\bibfnamefont{A.~D.} \bibnamefont{Stone}},
  \bibinfo{journal}{Phys. Rev. Lett.} \textbf{\bibinfo{volume}{55}},
  \bibinfo{pages}{1622} (\bibinfo{year}{1985}).

\bibitem[{\citenamefont{Altshuler}(1985)}]{altshuler1985}
\bibinfo{author}{\bibfnamefont{B.~L.} \bibnamefont{Altshuler}},
  \bibinfo{journal}{JETP Lett.} \textbf{\bibinfo{volume}{41}},
  \bibinfo{pages}{648} (\bibinfo{year}{1985}).

\bibitem[{\citenamefont{Lee et~al.}(1987)\citenamefont{Lee, Stone, and
  Fukuyama}}]{PhysRevB.35.1039}
\bibinfo{author}{\bibfnamefont{P.~A.} \bibnamefont{Lee}},
  \bibinfo{author}{\bibfnamefont{A.~D.} \bibnamefont{Stone}}, \bibnamefont{and}
  \bibinfo{author}{\bibfnamefont{H.}~\bibnamefont{Fukuyama}},
  \bibinfo{journal}{Phys. Rev. B} \textbf{\bibinfo{volume}{35}},
  \bibinfo{pages}{1039} (\bibinfo{year}{1987}).

\bibitem[{\citenamefont{Baibich et~al.}(1988)\citenamefont{Baibich, Broto,
  Fert, Van~Dau, Petroff, Etienne, Creuzet, Friederich, and Chazelas}}]{gmr1}
\bibinfo{author}{\bibfnamefont{M.~N.} \bibnamefont{Baibich}},
  \bibinfo{author}{\bibfnamefont{J.~M.} \bibnamefont{Broto}},
  \bibinfo{author}{\bibfnamefont{A.}~\bibnamefont{Fert}},
  \bibinfo{author}{\bibfnamefont{F.~N.} \bibnamefont{Van~Dau}},
  \bibinfo{author}{\bibfnamefont{F.}~\bibnamefont{Petroff}},
  \bibinfo{author}{\bibfnamefont{P.}~\bibnamefont{Etienne}},
  \bibinfo{author}{\bibfnamefont{G.}~\bibnamefont{Creuzet}},
  \bibinfo{author}{\bibfnamefont{A.}~\bibnamefont{Friederich}},
  \bibnamefont{and} \bibinfo{author}{\bibfnamefont{J.}~\bibnamefont{Chazelas}},
  \bibinfo{journal}{Phys. Rev. Lett.} \textbf{\bibinfo{volume}{61}},
  \bibinfo{pages}{2472} (\bibinfo{year}{1988}).

\bibitem[{\citenamefont{Binasch et~al.}(1989)\citenamefont{Binasch, Gr\"unberg,
  Saurenbach, and Zinn}}]{gmr2}
\bibinfo{author}{\bibfnamefont{G.}~\bibnamefont{Binasch}},
  \bibinfo{author}{\bibfnamefont{P.}~\bibnamefont{Gr\"unberg}},
  \bibinfo{author}{\bibfnamefont{F.}~\bibnamefont{Saurenbach}},
  \bibnamefont{and} \bibinfo{author}{\bibfnamefont{W.}~\bibnamefont{Zinn}},
  \bibinfo{journal}{Phys. Rev. B} \textbf{\bibinfo{volume}{39}},
  \bibinfo{pages}{4828} (\bibinfo{year}{1989}).

\bibitem[{\citenamefont{Zuti\'{c} et~al.}(2004)\citenamefont{Zuti\'{c}, Fabian,
  and {Das Sarma}}}]{spinrev}
\bibinfo{author}{\bibfnamefont{I.}~\bibnamefont{Zuti\'{c}}},
  \bibinfo{author}{\bibfnamefont{J.}~\bibnamefont{Fabian}}, \bibnamefont{and}
  \bibinfo{author}{\bibfnamefont{S.}~\bibnamefont{{Das Sarma}}},
  \bibinfo{journal}{Rev. Mod. Phys.} \textbf{\bibinfo{volume}{76}},
  \bibinfo{eid}{323} (pages~\bibinfo{numpages}{88}) (\bibinfo{year}{2004}).

\bibitem[{\citenamefont{Aprili et~al.}(1997)\citenamefont{Aprili, Lesueur,
  Dumoulin, and N\'{e}dellec}}]{kn:aprili1997}
\bibinfo{author}{\bibfnamefont{M.}~\bibnamefont{Aprili}},
  \bibinfo{author}{\bibfnamefont{J.}~\bibnamefont{Lesueur}},
  \bibinfo{author}{\bibfnamefont{L.}~\bibnamefont{Dumoulin}}, \bibnamefont{and}
  \bibinfo{author}{\bibfnamefont{P.}~\bibnamefont{N\'{e}dellec}},
  \bibinfo{journal}{Solid State Comm.} \textbf{\bibinfo{volume}{102}},
  \bibinfo{pages}{41} (\bibinfo{year}{1997}).

\bibitem[{\citenamefont{Wei et~al.}(2006)\citenamefont{Wei, Liu, Zhang, and
  Davidovi\'{c}}}]{kn:wei2006}
\bibinfo{author}{\bibfnamefont{Y.~G.} \bibnamefont{Wei}},
  \bibinfo{author}{\bibfnamefont{X.~Y.} \bibnamefont{Liu}},
  \bibinfo{author}{\bibfnamefont{L.~Y.} \bibnamefont{Zhang}}, \bibnamefont{and}
  \bibinfo{author}{\bibfnamefont{D.}~\bibnamefont{Davidovi\'{c}}},
  \bibinfo{journal}{Phys. Rev. Lett.} \textbf{\bibinfo{volume}{96}},
  \bibinfo{pages}{146803} (\bibinfo{year}{2006}).

\bibitem[{\citenamefont{Kasai et~al.}(2003)\citenamefont{Kasai, Saitoh, and
  Miyajima}}]{kn:kasai2003}
\bibinfo{author}{\bibfnamefont{S.}~\bibnamefont{Kasai}},
  \bibinfo{author}{\bibfnamefont{E.}~\bibnamefont{Saitoh}}, \bibnamefont{and}
  \bibinfo{author}{\bibfnamefont{H.}~\bibnamefont{Miyajima}},
  \bibinfo{journal}{Journal of Applied Physics} \textbf{\bibinfo{volume}{93}},
  \bibinfo{pages}{8427} (\bibinfo{year}{2003}).

\bibitem[{\citenamefont{Lee et~al.}(2007)\citenamefont{Lee, Trionfi,
  Schallenberg, Munekata, and Natelson}}]{kn:lee2007}
\bibinfo{author}{\bibfnamefont{S.}~\bibnamefont{Lee}},
  \bibinfo{author}{\bibfnamefont{A.}~\bibnamefont{Trionfi}},
  \bibinfo{author}{\bibfnamefont{T.}~\bibnamefont{Schallenberg}},
  \bibinfo{author}{\bibfnamefont{H.}~\bibnamefont{Munekata}}, \bibnamefont{and}
  \bibinfo{author}{\bibfnamefont{D.}~\bibnamefont{Natelson}},
  \bibinfo{journal}{Appl.\ Phys.\ Lett.} \textbf{\bibinfo{volume}{90}},
  \bibinfo{pages}{032105} (\bibinfo{year}{2007}).

\bibitem[{\citenamefont{Vila et~al.}(2007)\citenamefont{Vila, Giraud,
  Thevenard, Lema\^{i}tre, Pierre, Dufouleur, Mailly, Barbara, and
  Faini}}]{kn:vila2007}
\bibinfo{author}{\bibfnamefont{L.}~\bibnamefont{Vila}},
  \bibinfo{author}{\bibfnamefont{R.}~\bibnamefont{Giraud}},
  \bibinfo{author}{\bibfnamefont{L.}~\bibnamefont{Thevenard}},
  \bibinfo{author}{\bibfnamefont{A.}~\bibnamefont{Lema\^{i}tre}},
  \bibinfo{author}{\bibfnamefont{F.}~\bibnamefont{Pierre}},
  \bibinfo{author}{\bibfnamefont{J.}~\bibnamefont{Dufouleur}},
  \bibinfo{author}{\bibfnamefont{D.}~\bibnamefont{Mailly}},
  \bibinfo{author}{\bibfnamefont{B.}~\bibnamefont{Barbara}}, \bibnamefont{and}
  \bibinfo{author}{\bibfnamefont{G.}~\bibnamefont{Faini}},
  \bibinfo{journal}{Phys. Rev. Lett.} \textbf{\bibinfo{volume}{98}},
  \bibinfo{pages}{027204} (\bibinfo{year}{2007}).

\bibitem[{\citenamefont{Lee et~al.}(2004)\citenamefont{Lee, Trionfi, and
  Natelson}}]{kn:lee2004}
\bibinfo{author}{\bibfnamefont{S.}~\bibnamefont{Lee}},
  \bibinfo{author}{\bibfnamefont{A.}~\bibnamefont{Trionfi}}, \bibnamefont{and}
  \bibinfo{author}{\bibfnamefont{D.}~\bibnamefont{Natelson}},
  \bibinfo{journal}{Phys. Rev. B} \textbf{\bibinfo{volume}{70}},
  \bibinfo{pages}{212407} (\bibinfo{year}{2004}).

\bibitem[{\citenamefont{Bolotin et~al.}(2006)\citenamefont{Bolotin, Kuemmeth,
  and Ralph}}]{kn:bolotin2006}
\bibinfo{author}{\bibfnamefont{K.~I.} \bibnamefont{Bolotin}},
  \bibinfo{author}{\bibfnamefont{F.}~\bibnamefont{Kuemmeth}}, \bibnamefont{and}
  \bibinfo{author}{\bibfnamefont{D.~C.} \bibnamefont{Ralph}},
  \bibinfo{journal}{Phys. Rev. Lett.} \textbf{\bibinfo{volume}{97}},
  \bibinfo{pages}{127202} (\bibinfo{year}{2006}).

\bibitem[{\citenamefont{Neumaier et~al.}(2008)\citenamefont{Neumaier, Vogl,
  Eroms, and Weiss}}]{kn:neumaier2008}
\bibinfo{author}{\bibfnamefont{D.}~\bibnamefont{Neumaier}},
  \bibinfo{author}{\bibfnamefont{A.}~\bibnamefont{Vogl}},
  \bibinfo{author}{\bibfnamefont{J.}~\bibnamefont{Eroms}}, \bibnamefont{and}
  \bibinfo{author}{\bibfnamefont{D.}~\bibnamefont{Weiss}},
  \bibinfo{journal}{Phys. Rev. B} \textbf{\bibinfo{volume}{78}},
  \bibinfo{pages}{174424} (\bibinfo{year}{2008}).

\bibitem[{\citenamefont{Dugaev et~al.}(2001)\citenamefont{Dugaev, Bruno, and
  Barna\ifmmode~\acute{s}\else \'{s}\fi{}}}]{PhysRevB.64.144423}
\bibinfo{author}{\bibfnamefont{V.~K.} \bibnamefont{Dugaev}},
  \bibinfo{author}{\bibfnamefont{P.}~\bibnamefont{Bruno}}, \bibnamefont{and}
  \bibinfo{author}{\bibfnamefont{J.}~\bibnamefont{Barna\ifmmode~\acute{s}\else
  \'{s}\fi{}}}, \bibinfo{journal}{Phys. Rev. B} \textbf{\bibinfo{volume}{64}},
  \bibinfo{pages}{144423} (\bibinfo{year}{2001}).

\bibitem[{\citenamefont{Takane and Koyama}(2000)}]{takane2}
\bibinfo{author}{\bibfnamefont{Y.}~\bibnamefont{Takane}} \bibnamefont{and}
  \bibinfo{author}{\bibfnamefont{Y.}~\bibnamefont{Koyama}},
  \bibinfo{journal}{J. Phys. Soc. Jpn.} \textbf{\bibinfo{volume}{69}},
  \bibinfo{pages}{328} (\bibinfo{year}{2000}).

\bibitem[{\citenamefont{Tatara and Fukuyama}(1997)}]{PhysRevLett.78.3773}
\bibinfo{author}{\bibfnamefont{G.}~\bibnamefont{Tatara}} \bibnamefont{and}
  \bibinfo{author}{\bibfnamefont{H.}~\bibnamefont{Fukuyama}},
  \bibinfo{journal}{Phys. Rev. Lett.} \textbf{\bibinfo{volume}{78}},
  \bibinfo{pages}{3773} (\bibinfo{year}{1997}).

\bibitem[{\citenamefont{Lyanda-Geller et~al.}(1998)\citenamefont{Lyanda-Geller,
  Aleiner, and Goldbart}}]{PhysRevLett.81.3215}
\bibinfo{author}{\bibfnamefont{Y.}~\bibnamefont{Lyanda-Geller}},
  \bibinfo{author}{\bibfnamefont{I.~L.} \bibnamefont{Aleiner}},
  \bibnamefont{and} \bibinfo{author}{\bibfnamefont{P.~M.}
  \bibnamefont{Goldbart}}, \bibinfo{journal}{Phys. Rev. Lett.}
  \textbf{\bibinfo{volume}{81}}, \bibinfo{pages}{3215} (\bibinfo{year}{1998}).

\bibitem[{\citenamefont{Adam et~al.}(2006)\citenamefont{Adam, Kindermann,
  Rahav, and Brouwer}}]{kn:adam2006b}
\bibinfo{author}{\bibfnamefont{S.}~\bibnamefont{Adam}},
  \bibinfo{author}{\bibfnamefont{M.}~\bibnamefont{Kindermann}},
  \bibinfo{author}{\bibfnamefont{S.}~\bibnamefont{Rahav}}, \bibnamefont{and}
  \bibinfo{author}{\bibfnamefont{P.~W.} \bibnamefont{Brouwer}},
  \bibinfo{journal}{Phys. Rev. B} \textbf{\bibinfo{volume}{73}},
  \bibinfo{pages}{212408} (\bibinfo{year}{2006}).

\bibitem[{\citenamefont{Takane}(2003)}]{JPSJ.72.1155}
\bibinfo{author}{\bibfnamefont{Y.}~\bibnamefont{Takane}}, \bibinfo{journal}{J.
  Phys. Soc. Jpn.} \textbf{\bibinfo{volume}{72}}, \bibinfo{pages}{1155}
  (\bibinfo{year}{2003}).

\bibitem[{\citenamefont{W\"{o}lfle and Muttalib}(2010)}]{muttalibwoelfle}
\bibinfo{author}{\bibfnamefont{P.}~\bibnamefont{W\"{o}lfle}} \bibnamefont{and}
  \bibinfo{author}{\bibfnamefont{K.~A.} \bibnamefont{Muttalib}}, in
  \emph{\bibinfo{booktitle}{Perspectives of mesoscopic physics}}, edited by
  \bibinfo{editor}{\bibfnamefont{A.}~\bibnamefont{Aharony}} \bibnamefont{and}
  \bibinfo{editor}{\bibfnamefont{O.}~\bibnamefont{{Entin-Wohlman}}}
  (\bibinfo{publisher}{World Scientific}, \bibinfo{year}{2010}).

\bibitem[{\citenamefont{Izuyama et~al.}(1963)\citenamefont{Izuyama, Kim, and
  Kubo}}]{kubospinsus}
\bibinfo{author}{\bibfnamefont{T.}~\bibnamefont{Izuyama}},
  \bibinfo{author}{\bibfnamefont{D.-J.} \bibnamefont{Kim}}, \bibnamefont{and}
  \bibinfo{author}{\bibfnamefont{R.}~\bibnamefont{Kubo}}, \bibinfo{journal}{J.
  Phys. Soc. Jpn.} \textbf{\bibinfo{volume}{18}}, \bibinfo{pages}{1025}
  (\bibinfo{year}{1963}).

\bibitem[{\citenamefont{Landau and Lifshitz}(1980)}]{ll5}
\bibinfo{author}{\bibfnamefont{L.~D.} \bibnamefont{Landau}} \bibnamefont{and}
  \bibinfo{author}{\bibfnamefont{E.~M.} \bibnamefont{Lifshitz}},
  \emph{\bibinfo{title}{Statistical Physics, Part 1}}
  (\bibinfo{publisher}{Pergamon Press}, \bibinfo{year}{1980}).

\bibitem[{\citenamefont{Kittel}(1963)}]{kittel}
\bibinfo{author}{\bibfnamefont{C.}~\bibnamefont{Kittel}},
  \emph{\bibinfo{title}{Quantum Theory of Solids}} (\bibinfo{publisher}{Wiley,
  New York}, \bibinfo{year}{1963}).

\bibitem[{\citenamefont{{O'Handley}}(2000)}]{mmm}
\bibinfo{author}{\bibfnamefont{R.~C.} \bibnamefont{{O'Handley}}},
  \emph{\bibinfo{title}{Modern Magnetic Materials}} (\bibinfo{publisher}{Wiley,
  New York}, \bibinfo{year}{2000}).

\bibitem[{\citenamefont{Raquet et~al.}(2002)\citenamefont{Raquet, Viret,
  Sondergard, Cespedes, and Mamy}}]{PhysRevB.66.024433}
\bibinfo{author}{\bibfnamefont{B.}~\bibnamefont{Raquet}},
  \bibinfo{author}{\bibfnamefont{M.}~\bibnamefont{Viret}},
  \bibinfo{author}{\bibfnamefont{E.}~\bibnamefont{Sondergard}},
  \bibinfo{author}{\bibfnamefont{O.}~\bibnamefont{Cespedes}}, \bibnamefont{and}
  \bibinfo{author}{\bibfnamefont{R.}~\bibnamefont{Mamy}},
  \bibinfo{journal}{Phys. Rev. B} \textbf{\bibinfo{volume}{66}},
  \bibinfo{pages}{024433} (\bibinfo{year}{2002}).

\bibitem[{\citenamefont{Ashcroft and Mermin}(1974)}]{ashmer}
\bibinfo{author}{\bibfnamefont{N.~W.} \bibnamefont{Ashcroft}} \bibnamefont{and}
  \bibinfo{author}{\bibfnamefont{N.~D.} \bibnamefont{Mermin}},
  \emph{\bibinfo{title}{Solid State Physics}} (\bibinfo{publisher}{Saunders,
  New York}, \bibinfo{year}{1974}).

\bibitem[{\citenamefont{Birge et~al.}(1989)\citenamefont{Birge, Golding, and
  Haemmerle}}]{PhysRevLett.62.195}
\bibinfo{author}{\bibfnamefont{N.~O.} \bibnamefont{Birge}},
  \bibinfo{author}{\bibfnamefont{B.}~\bibnamefont{Golding}}, \bibnamefont{and}
  \bibinfo{author}{\bibfnamefont{W.~H.} \bibnamefont{Haemmerle}},
  \bibinfo{journal}{Phys. Rev. Lett.} \textbf{\bibinfo{volume}{62}},
  \bibinfo{pages}{195} (\bibinfo{year}{1989}).

\bibitem[{\citenamefont{Aleiner and Blanter}(2002)}]{PhysRevB.65.115317}
\bibinfo{author}{\bibfnamefont{I.~L.} \bibnamefont{Aleiner}} \bibnamefont{and}
  \bibinfo{author}{\bibfnamefont{Y.~M.} \bibnamefont{Blanter}},
  \bibinfo{journal}{Phys. Rev. B} \textbf{\bibinfo{volume}{65}},
  \bibinfo{pages}{115317} (\bibinfo{year}{2002}).

\end{thebibliography}
\end{document}